\documentclass[english,aps,superscriptaddress,showkeys,showpacs,prepri]{revtex4}
\usepackage[T1]{fontenc}
\usepackage[latin9]{inputenc}
\usepackage[letterpaper]{geometry}
\usepackage{units}
\usepackage{bbding}
\usepackage{amsmath}
\usepackage{amssymb}
\usepackage{graphicx}
\usepackage{esint}

\makeatletter

\providecommand{\tabularnewline}{\\}

\@ifundefined{textcolor}{}
{%
 \definecolor{BLACK}{gray}{0}
 \definecolor{WHITE}{gray}{1}
 \definecolor{RED}{rgb}{1,0,0}
 \definecolor{GREEN}{rgb}{0,1,0}
 \definecolor{BLUE}{rgb}{0,0,1}
 \definecolor{CYAN}{cmyk}{1,0,0,0}
 \definecolor{MAGENTA}{cmyk}{0,1,0,0}
 \definecolor{YELLOW}{cmyk}{0,0,1,0}
}

\usepackage{doi}
\usepackage{hyperref}

\makeatother

\usepackage{babel}
\begin{document}

\title{A Parametric Study of Extended-MHD Drift Tearing}

\author{J. R. King}

\affiliation{Tech-X Corporation, 5621 Arapahoe Ave. Suite A Boulder, CO 80303}

\author{S. E. Kruger}

\affiliation{Tech-X Corporation, 5621 Arapahoe Ave. Suite A Boulder, CO 80303}

\date{draft \today}
\begin{abstract}
The linear drift-tearing mode is analyzed for different regimes of
the plasma-$\beta$, ion-skin-depth parameter space with an unreduced,
extended-MHD model. New dispersion relations are found at moderate
plasma $\beta$ and previous drift-tearing results are classified
as applicable at small plasma $\beta$. The drift stabilization of
the mode in the regimes varies from non-existent/weak to complete.
As the diamagnetic-drift frequency is proportional to the plasma $\beta$,
verification exercises with unreduced, extended-MHD models in the
small plasma-$\beta$ regimes are impractical. The new dispersion
relations in the moderate plasma-$\beta$ regimes are used to verify
the extended-MHD implementation of the NIMROD code {[}C. R. Sovinec
\textit{et al}., J. Comput. Phys. \textbf{195}, 355 (2004){]}. Given
the small boundary-layer skin depth, discussion of the validity of
the first-order finite-Larmour-radius model is presented.
\end{abstract}

\keywords{Drift-tearing mode, Extended-MHD modeling, Code Verification}

\pacs{52.30.-q, 52.65.-y, 52.35.-g, 52.55.Tn, 52.40.Hf, 02.60.Lj, 52.35.Vd}

\maketitle

\section{Introduction \label{sec:Intro}}

\noindent Experimental, fusion-plasma discharges typically operate
in regimes away from ideal-magnetohydrodynamic (MHD) stability boundaries.
The ideal-MHD modes that exist outside these boundaries, which are
unable to modify the magnetic topology, are often deleterious to confinement
and can lead to a rapid loss of the plasma stored energy. Analysis
with a resistive-MHD model shows a second class of modes are possible.
These resistive-MHD modes are a combination of macroscopic ideal-MHD
behavior through-out most of the plasma volume and boundary-layer
dynamics where resistivity is important near a resonant magnetic-flux
surface, a surface where the mode structure and the magnetic topology
are aligned in poloidal and toroidal periodic variation. Although
the plasma dynamics associated with these resistive modes are usually
less violent than ideal modes, finite resistivity allows for modification
of the magnetic topology. For example, magnetic islands formed from
saturated resistive-tearing modes can enhance energy and particle
transport from the plasma core to the edge via large field-aligned
transport. 

The tearing instability \cite{Furth_63_PoF_459} is one such multi-scale
mode: a combination of macroscopic structure, the ideal-MHD response
through-out most of the plasma volume; and microscopic structure,
the boundary-layer physics near the resonant surface which minimally
includes resistive MHD. Ideal-MHD flows advect magnetic flux to the
resonant surface where a large, localized current sheet is formed.
This leads to slow growth on a hybrid-time scale that is a combination
of the ideal Alfvén time and the time scale of the pertinent boundary-layer
physics. With a resistive-MHD model, the current-sheet size is determined
by the magnitude of the plasma resistivity: smaller resistivity results
in a more localized layer. In high-temperature fusion plasmas, which
have very small resistivity, the boundary-layer width can approach
the ion gyroradius where finite-Larmour-radius (FLR) and electron-ion-fluid-decoupling
effects become important. When the more mobile electron fluid is decoupled
from the ion fluid near the layer, it can more effectively transport
flux into the layer and thus destabilize the mode (increase the growth
rate). Alternatively, when the fluids are decoupled and drift in opposed
directions within the resonant flux surface, the sheared relative
motion can stabilize the mode (reduce the growth rate). A sufficient
model to capture these FLR effects to first order is extended-MHD
with Braginskii-like closures \cite{Braginskii_65__,Catto_04_PoP_90,Ramos_2010_PoP1,Ramos_2011_PoP1}.
The zeroth-order plasma drift, the $\mathbf{E}\times\mathbf{B}$ drift,
causes the electron and ion fluids to drift with the same velocity
and thus are not stabilizing. The first-order FLR drifts have a orientation
that is dependent on the sign of the charge of the species and thus
are stabilizing. With respect to influence on the tearing mode, the
most studied first-order FLR drift is the fluid diamagnetic drift
\cite{Coppi_64_PoF_1501} but stabilizing effects are also attributed
to drifts proportional to the gradient and curvature of the magnetic
field \cite{King_11_PoP_42303}. 

\begin{center}
\begin{figure}
\centering{}\includegraphics[width=0.45\paperwidth]{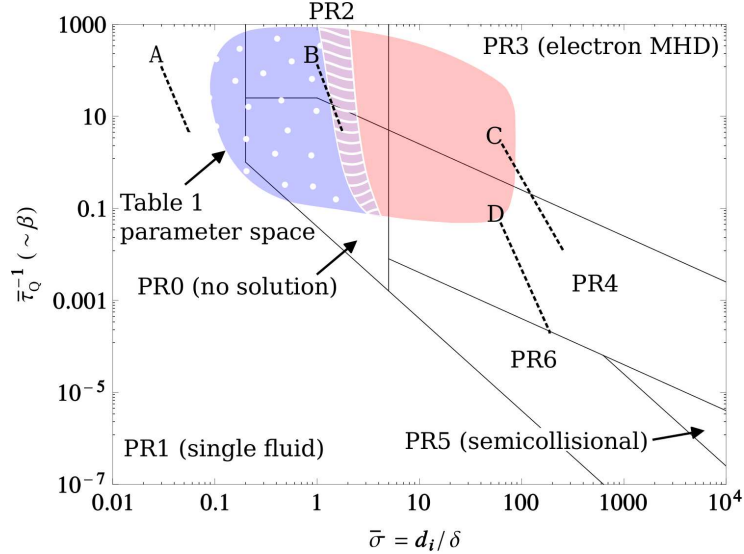}\caption{Tearing mode parameter space in terms of normalized $\beta$, $\bar{\tau}_{Q}^{-1}$,
and $d_{i}$, $\bar{\sigma}$, (as originally defined in Ref.~\cite{Ahedo_09_PPaCF_55018}).
Growth rates from PR1 through PR5 are used within the normalizations
as appropriate. The colored (dark) area maps the region of interest
for tokamak fusion plasmas with parameter ranges as defined in Tab.~\ref{tab:experimentalParam}.
A first-order ion-FLR model is valid in the blue, dotted region ($\rho_{i}<0.25\delta$),
and invalid in the solid, red region ($\rho_{i}>0.25\delta$). There
is a wavy, purple region which contains both valid and invalid cases
as the normalized parameter space and $\rho_{i}/\delta$ do not have
a one-to-one mapping. The diagram uses a small parameter value of
$0.04$ to determine regime boundaries. Points A through D correspond
to the $\omega_{*}\rightarrow0$ limit of the verification scans of
Sec.~\ref{sec:Verification} and the modification of $\bar{\tau}$
and $\bar{\sigma}$ as $\omega_{*}$ is increased is illustrated with
dashed lines for the ranges of $\omega_{*}$ included in the verification
exercises. \label{fig:paramSpace}}
\end{figure}

\par\end{center}

A previous parametric regime analysis of the tearing mode without
drift effects is given by Ahedo and Ramos \cite{Ahedo_09_PPaCF_55018}.
They characterize small-$\Delta^{\prime}$ tearing-mode parameter
space by seven regimes as illustrated schematically in Fig.~\ref{fig:paramSpace}.
A single-fluid model (resistive MHD) describes the dynamics in parameter-space-region
PR1 as first discovered by Furth et al.~\cite{Furth_63_PoF_459}.
In PR5, at very small values of $\beta$ and large ion-skin depth
($d_{i}$), the semicollisional description of Drake and Lee is valid
\cite{Drake_77_PoF_1341}. Without drift-effects, the semicollisional
description is valid when $\beta$ is smaller than the square of the
tearing skin depth normalized by the mode wavenumber (ignoring some
factors of order unity). Thus even for a large tearing skin depth
of 1cm, the validity constraint is still approximately $\beta<<10^{-4}$
with a mode wavelength of 1m and it is unlikely this regime is relevant
to tokamak discharges. At moderate values of the plasma-$\beta$ parameter
and $d_{i}$ (thus moderate values of the ion gyroradius, $\rho_{i}\sim\sqrt{\beta}d_{i}$)
the tearing dispersion relation from electron-MHD \cite{Bulanov_92_PoFB_2499}
is recovered (PR3). Mirnov et al. derive a unified dispersion relation
for PR4 which limits to that found in PR3 and PR5 \cite{Mirnov_04_PoP_4468}.
They describe the decoupling effects of the mode as mediated through
interaction with the kinetic-Alfvén wave in PR3 and the whistler wave
in PR5. The dispersion relations for the remaining transitional regimes
in this parameter space (PR2 and PR6) are derived by Ahedo and Ramos
\cite{Ahedo_09_PPaCF_55018}. There is no known solution in PR0.

Our results largely follow the parameter-space characterization of
Ref.~\cite{Ahedo_09_PPaCF_55018}, however our calculations include
diamagnetic and magnetic-field-gradient drift contributions. In a
sense, the main concept of our study is to add a third dimension out
of the page of Fig.~\ref{fig:paramSpace} that corresponds to the
drift frequency. In Sec.~\ref{sec:ModelEqnsOrdering}, we describe
the extended-MHD model and our small-$\Delta^{\prime}$, large-guide-field
assumptions. These equations are linearized and reduced to a system
of two second-order equations in Secs. \ref{sec:LinEqns} and \ref{sec:SysEqns}.
Our intention is to clarify the relevant regimes to fusion plasmas
and benchmark extended-MHD drift-tearing computations which use an
unreduced-MHD model. As such, our study differs from much of the prior
work in that we do not start with a reduced-MHD model, but rather
we apply tearing ordering to the full extended-MHD equations. Our
main dispersion relation results are derived in Sec.~\ref{sec:Dispersion}
for drift tearing in PR1 through PR5. We recover the result of Coppi
at small values of the plasma-$\beta$ parameter in the single-fluid
regime (PR1, Refs.~\cite{Coppi_64_PoF_1501,Biskamp_78_NF_1059})
and the result of Drake and Lee in the semicollisional regime (PR5,
Ref.~\cite{Drake_77_PoF_1341}). New dispersion relations are found
in PR2 through PR4. 

The linear drift-tearing response in the moderate-$\beta$ regimes
is necessary to verify extended-MHD codes at the parameters of typical
use cases. In Sec.~\ref{sec:Verification} we present the results
of a verification exercise between the NIMROD extended-MHD code \cite{Sovinec_04_JoCP_355}
and our new drift-tearing dispersion relations in PR2 through PR4.
The extended-MHD drift terms are complicated even in primitive form,
as described in Sec.~\ref{sec:ModelEqnsOrdering}. The complexity
of the model makes implementation and numerical analysis challenging
and thus makes verification all the more important. The NIMROD extended-MHD
algorithm has previously been benchmarked in PR1 through PR5 without
drift effects \cite{Sovinec_10_JoCP_5803}. Our verification exercise
with drift effects extends this benchmark to more experimentally relevant
regimes.

Our new verification scans of the drift frequency begin at points
A-D in the $\bar{\tau}_{Q}$-$\bar{\sigma}$ parameter space of Fig.~\ref{fig:paramSpace}.
As the drift frequency increases, the $\bar{\tau}_{Q}$ parameter
of Fig.~\ref{fig:paramSpace} increases and the single-fluid tearing
skin depth, $\delta$, decreases. Thus cases move within the $\bar{\tau}_{Q}$-$\bar{\sigma}$
parameter space of the figure down and slightly to the right as illustrated
in the figure by the dashed lines. As the diamagnetic-drift frequency
scales proportionally to $\beta$, we have found it difficult to satisfy
all of the analytic asymptotic-limit requirements of the tearing mode
while running with appreciable drift frequency for a verification
exercise purely in the low-$\beta$ semicollisional regime (PR5). 

\begin{table}
\begin{centering}
\begin{tabular}{cc}
parameter & range\tabularnewline
\hline 
\hline 
$S$ & $10^{7}$-$10^{9}$\tabularnewline
$k_{\perp}d_{i}$ & $0.01$-$1.0$\tabularnewline
$\beta$ & $0.005$-$0.1$\tabularnewline
$k_{\perp}^{-1}\Delta^{\prime}$ & 0.5-20\tabularnewline
$\epsilon_{B}$  & 0.02-0.5\tabularnewline
\end{tabular}
\par\end{centering}

\caption{Expected range of parameters for modern-tokamak-core experimental
conditions. The parameters are the Lundquist number, $S$, the ion
skin depth, $d_{i}$, the plasma $\beta$, the tearing stability parameter,
$\Delta^{\prime}$, and the guide-to-sheared magnetic-field ratio,
$\epsilon_{B}$ (definitions are provided in the text).\label{tab:experimentalParam} }
\end{table}

If the effect of electron inertia is larger than that of resistivity,
the dynamics are described by so-called collisionless physics. In
this regime, the growth rate is independent of resistivity. Electron
inertia scales proportionally to the electron skin depth squared,
where the ion and electron skin depths ($d_{i}$ and $d_{e}$) have
a fixed ratio equal to the square root of the mass ratio. Thus in
PR3 through PR5 as $d_{i}$ is increased in Fig.~\ref{fig:paramSpace},
the mode will ultimately become collisionless. Also for this reason
it is difficult, if not impossible, to compose collisionless cases
in PR6, PR1 and PR2 unless one is using a model with an enhanced electron
mass or operating with extremely low plasma $\beta$. We interpret
our drift-tearing results in the transition-to-collisionless electron-MHD
regime (PR3) and discuss implications for extended-MHD modeling in
Sec.~\ref{sec:Verification}. 

Fitzpatrick points out that when one compares the size of the electron
gyroradius, $\rho_{e}$, to the single-fluid tearing skin depth, $\delta$,
in the moderate-$\beta$ regimes (PR3-5) for collisionless cases,
a first-order electron-FLR model is invalid throughout all of the
electron-MHD regime (PR3) and much of PR4 as $\rho_{e}>\delta$. However
the model is valid in the semicollisional regime (PR5) \cite{Fitzpatrick_2010_PoP}.
A corollary to this argument is that a first-order ion-FLR model will
be invalid when $\rho_{i}>\delta$. Consider the expected fusion-plasma
parameters as listed in Table \ref{tab:experimentalParam}; the limits
in the $\bar{\tau}_{Q}$-$\bar{\sigma}$ parameter space defined by
the Tab.~\ref{tab:experimentalParam} parameters are superimposed
onto Fig.~\ref{fig:paramSpace}. A first-order ion-FLR model is valid
when $\sqrt{\beta}d_{i}\sim\rho_{i}<<\delta$, which encompasses many
of the fusion-relevant cases in PR0, PR1 and PR2. For these parameters,
a first-order electron-FLR model is always valid (with a realistic
mass ratio, $\mu=m_{e}/m_{i})$ as $\rho_{e}/\delta=\sqrt{\mu}\rho_{i}/\delta\sim\sqrt{\mu\beta}\bar{\sigma}$
and $\bar{\sigma}$ never exceeds a value of 100 with the parameters
of Tab.~\ref{tab:experimentalParam}. We note two reasons for studying
drift tearing with a first-order ion-FLR model outside its regime
of strict validity. First, the model may be outside the region of
strict validity only for \emph{linear} modes. With nonlinear dynamics,
the tearing skin depth is no longer a well defined concept and, strictly
from the linear definition, it broadens as the mode approaches saturation.
In these nonlinear regimes, model validity is determined largely by
the constraint $k\rho_{i}<<1$ that is more easily satisfied by long-wavelength
tearing instabilities ($k$ is the perturbation wavenumber). First-order
FLR, extended-MHD modeling is typically interested in the nonlinear
evolution of the plasma; however, most computations first encounter
a linear growth phase that is still important to both understand and
ensure that it is calculated correctly. Second, the mode dynamics
transition to an electron-MHD description as $\rho_{i}$ becomes large
and the ion fluid becomes demagnetized on the small tearing-skin-depth
scale and decoupled from the electron fluid. If a first-order FLR
model is capable of correctly modeling these electron-fluid dynamics,
it may be qualitatively descriptive of the electron dynamics outside
its regime of strict validity. Qualitatively descriptive but computationally
tractable first-order FLR, extended-MHD modeling is preferable to
modeling with full-orbit ion dynamics when the latter is computational
intractable.

\section{Model Equations and Orderings\label{sec:ModelEqnsOrdering}}

\noindent With an unreduced-MHD model, the plasma fluid is described
by a continuity equation,
\begin{equation}
\frac{\partial n}{\partial t}=-\nabla\cdot n\mathbf{v}\;,\label{eq:cont}
\end{equation}
for the plasma density ($n$) evolution, a center-of-mass momentum
equation, 
\begin{equation}
m_{i}n\frac{d\mathbf{v}}{dt}=\mathbf{J}\times\mathbf{B}-\nabla p-\nabla\cdot\mathbf{\mathbf{\Pi}}_{i}\;,\label{eq:mom}
\end{equation}
for the bulk-plasma velocity ($\mathbf{v}$), and an energy equation,
\begin{equation}
\frac{n}{\Gamma-1}\frac{d^{\alpha}T_{\alpha}}{dt}=-p_{\alpha}\nabla\cdot\mathbf{v}_{\alpha}-\nabla\cdot\mathbf{q}_{\alpha}\;,\label{eq:energy}
\end{equation}
for the plasma temperature ($T_{\alpha}$). The subscript indicates
either the ion or electron species, $m_{\alpha}$ is a species' mass,
and $\Gamma$ is the adiabatic index. The plasma is assumed to be
an ideal gas and thus the species pressure ($p_{\alpha}$; $p=\sum p_{\alpha}$)
is given by the ideal-gas law, $p_{\alpha}=nT_{\alpha}$. As appropriate
for low-frequency plasma dynamics, we assume quasi-neutrality ($n_{e}\simeq n_{i}$
for an ion charge state of unity) and drop the displacement-current
term in Ampere's law ($\mu_{0}\mathbf{J}=\nabla\times\mathbf{B}$
where $\mu_{0}$ is permeability of free space), which provides a
relation between the magnetic field ($\mathbf{B}$) and the current
density ($\mathbf{J}=ne(\mathbf{v}_{i}-\mathbf{v}_{e})$ where $e$
is the electron charge). These approximations analytically eliminate
both light and Langmuir waves. The electron momentum equation is used
as an expression for the electric field ($\mathbf{E}$),

\begin{equation}
\mathbf{E}=-\mathbf{v}\times\mathbf{B}+\frac{\mathbf{J}\times\mathbf{B}}{ne}-\frac{\nabla p_{e}}{ne}-\frac{\nabla\cdot\mathbf{\mathbf{\Pi}}_{e}}{ne}+\eta\mathbf{J}-\frac{m_{e}}{e}\frac{d^{e}\mathbf{v}_{e}}{dt}\;,\label{eq:genOhmsLaw}
\end{equation}
commonly referred to as the generalized Ohm's law ($m_{e}$ is the
electron mass and $\eta$ is the electrical resistivity caused by
electron-ion collisions). Faraday's law ($\partial\mathbf{B}/\partial t=-\nabla\times\mathbf{E}$)
in conjunction with Eqn.~(\ref{eq:genOhmsLaw}) produces the induction
equation, which describes the evolution of the magnetic field. This
system of equations is considered to be a two-fluid model when the
Hall term ($\mathbf{J}\times\mathbf{B}/ne$) is retained as the magnetic
field is then advected by the electron flow ($\mathbf{v}_{e}=\mathbf{v}_{i}-\mathbf{J}/ne$)
instead of bulk-flow advection from the $\mathbf{v}\times\mathbf{B}$
term. 

These equations require closure expressions for the stress tensors
($\mathbf{\Pi}_{\alpha}$) and heat fluxes ($\mathbf{q}_{\alpha}$).
We use the Braginskii-like \cite{Catto_04_PoP_90,Ramos_2010_PoP1,Ramos_2011_PoP1}
`cross' terms (first-order FLR terms) as the closure: gyroviscosity,
\begin{equation}
\mathbf{\Pi}_{\alpha}=\frac{m_{\alpha}p_{\alpha}}{4q_{\alpha}B}\left[\hat{\mathbf{b}}\times\mathbf{W}_{\alpha}\cdot\left(\mathbf{I}+3\hat{\mathbf{b}}\hat{\mathbf{b}}\right)-\left(\mathbf{I}+3\hat{\mathbf{b}}\hat{\mathbf{b}}\right)\cdot\mathbf{W}_{\alpha}\times\hat{\mathbf{b}}\right]\;,\label{eq:gyrovisc}
\end{equation}
and cross-heat flux,
\begin{equation}
\mathbf{q}=\frac{5p_{\alpha}}{2q_{\alpha}B_{0}}\hat{\mathbf{b}}\times\nabla T_{\alpha}\;,\label{eq:crossHeat}
\end{equation}
where $q_{\alpha}$ is a species' charge. The rate-of-strain tensor
($\mathbf{W}_{\alpha}$) is defined as $\mathbf{W}_{\alpha}=\nabla\mathbf{v}_{\alpha}+\nabla\mathbf{v}_{\alpha}^{T}-(2/3)\mathbf{I}\nabla\cdot\mathbf{v}_{\alpha}$.
This choice of closure neglects the perpendicular and parallel (to
$\mathbf{B}$) closure terms and additional contributions to the gyroviscous
stress \cite{Catto_04_PoP_90,Ramos_05_PoP_112301}; however, the retained
terms are commonly included in state-of-the-art extended-MHD codes
and have contributions that enter the model equations on the same
order as the diamagnetic-drift terms. 

To further estimate the importance of the cross-closure terms, consider
flows on the order of the sound speed, $c_{s}=\sqrt{\Gamma\left(T_{i}+T_{e}\right)/m_{i}}$,
which for comparable species' temperatures is on the same order as
the ion thermal speed, $v_{T\alpha}=\sqrt{T_{\alpha}/m_{\alpha}}$.
The ion gyroviscous term then scales as $\rho_{i}/L$ relative to
the $\nabla p$ term in the momentum equation, Eqn.~(\ref{eq:mom}),
whereas the electron gyroviscous term scales as $\sqrt{m_{e}/m_{i}}\left(\rho_{e}/L\right)$
relative to the $\nabla p_{e}$ term in the generalized Ohm's law,
Eqn.~(\ref{eq:genOhmsLaw}). Here $\rho_{\alpha}=v_{T\alpha}/\omega_{c\alpha}$
is the gyroradius where $\omega_{c\alpha}=q_{\alpha}B/m_{\alpha}$
is the gyrofrequency and $L$ is a characteristic gradient length
scale. Furthermore, the ratio of the electron to ion gyroradius is
the square root of the mass ratio, $\sqrt{m_{e}/m_{i}}$. Thus if
the ion gyroviscous term is significant and the first-order ion-FLR
model remains valid, $\rho_{i}/L\lesssim\mathcal{O}\left(1\right)$,
then the electron gyroviscous term is expected to be smaller than
other terms in the generalized Ohm's law by at least the mass ratio.
As such, we neglect contributions from electron gyroviscosity in our
equations. This assumption leads to a break-down of the model for
the collisionless drift-tearing mode where these scalings do not apply
within the layer, as discussed in Sec.~\ref{sec:Verification}. Next
consider the cross-heat flux terms relative to $p_{\alpha}\nabla\cdot\mathbf{v}_{\alpha}$
in Eqn.~(\ref{eq:energy}). The ion cross-heat flux scales as $\rho_{i}/L$,
but the electron cross-heat flux scales as $\sqrt{m_{i}/m_{e}}\left(\rho_{e}/L\right)$.
Thus if the ion cross-heat flux is significant ($\rho_{i}/L\lesssim\mathcal{O}\left(1\right)$)
then the electron cross-heat flux enters the equations on the same
order and must be retained.  

For the purposes of our study, the tearing instability is generated
from an imposed $\hat{z}$-oriented current sheet in a Cartesian slab.
There are distant conducting walls at $x=\pm\infty$, the $\hat{y}$
and $\hat{z}$ directions are infinite, and the $\hat{z}$ direction
is symmetric. The tearing mode drive is fueled by free energy from
the global configuration but growth the of the mode is limited by
the small-scale physics that breaks the frozen-flux theorem within
the tearing boundary layer. As this boundary-layer physics is the
focus of our study, the slab configuration is locally analogous to
a toroidal configuration without curvature contributions where $\hat{x}$
is a radial (flux) coordinate, $\hat{y}$ is approximately a cross-field
coordinate and $\hat{z}$ is approximately a parallel-field coordinate.
We decompose all fields into imposed, $x$-dependent, background fields
(`0' subscript) and periodic-in-$\hat{y}$, perturbation fields (tilde),
e.g. $\mathbf{B}=\mathbf{B}_{0}\left(x\right)+\tilde{\mathbf{B}}\left(x\right)\exp\left(iky+\gamma t\right)$.
Here $\mathbf{k}=k\hat{y}$ is the perturbation wavenumber and $\gamma$
is the complex growth rate. The radial ($\hat{x}$) component of all
background vector fields is zero. Perturbation vector fields and wavenumber
use a magnetic-coordinate system where the $\hat{y}$ and $\hat{z}$
components are expressed as parallel-to and perpendicular-to the magnetic
field.

In our subsequent analysis, we ignore the effects of flow shear but
retain the effect of advection by bulk background flows. We impose
orderings appropriate for the tearing boundary layer: (1) the equilibrium
magnetic-shear-length scale ($L_{s}$) is comparable to the inverse
wavelength, $kL_{s}\sim\mathcal{O}\left(1\right)$; (2) a moderately
large guide- to shear-magnetic-field ratio, such that $\epsilon_{B}=B_{z}\left(x=0\right)/B_{y}\left(x=\infty\right)\sim\mathcal{O}\left(\epsilon^{3/4}\right)$;
(3) a small tearing skin depth ($\delta$), $kx\sim k\delta\sim\mathcal{O}\left(\epsilon\right)$;
(4) slow dynamics, $\omega\tau_{A}\sim\mathcal{O}\left(\epsilon^{3/2}\right)$;
and (5) slowly varying profiles within the layer, e.g.~$\delta n_{0}^{\prime}/n_{0}\sim\mathcal{O}\left(\epsilon\right)$.
Here $\epsilon$ is a small parameter ($\epsilon<<1$) and $\tau_{A}^{-1}=kv_{A}=kB_{0}/\sqrt{m_{i}n_{0}\mu_{0}}$
is the Alfvén time. 

These assumptions are consistent with the expected conditions for
core tearing in a high-temperature tokamak discharge. Our analysis
can accommodate very small values of $\beta$ ($c_{s}^{2}/v_{A}^{2}$),
however we assume the growth rate is subsonic ($\gamma^{2}<<k^{2}c_{s}^{2}$).
Ahedo and Ramos show that when this assumption is violated without
drift effects, the eigenfunction structure is modified but the growth
rate is unchanged \cite{Ahedo_2012_PoP1}. We assume that the electron-inertia
term is dominated by the contribution from current density and that
advection in the electron inertia term, which is of the same order
as electron gyroviscosity, is small. Thus after linearization, 
\begin{equation}
\frac{m_{e}}{e}\frac{d^{e}\mathbf{v}_{e}}{dt}\simeq\frac{m_{e}}{e}\left(\mathbf{v}_{e0}\cdot\nabla+\frac{\partial}{\partial t}\right)\tilde{\mathbf{v}}_{e}\simeq\frac{m_{e}}{n_{e}e^{2}}\frac{\partial\tilde{\mathbf{J}}}{\partial t}\;.
\end{equation}
As the linearized contributions from both electron inertia and resistivity
are proportional to $\tilde{\mathbf{J}}$, we simplify the subsequent
equations by combining these terms and forming a generalized resistivity,
\begin{equation}
\frac{\eta_{g}}{\mu_{0}}=\frac{\eta}{\mu_{0}}+d_{e}^{2}\gamma\;.
\end{equation}
The relative magnitude of resistivity compared to electron inertia
classifies the tearing mode as collisionless ($d_{e}^{2}\gamma>>\eta/\mu_{0}$)
or collisional ($d_{e}^{2}\gamma<<\eta/\mu_{0}$) where $d_{\alpha}$
is a species' skin depth ($\sqrt{m_{\alpha}/\mu_{0}ne^{2}}$). Similarly,
we use a generalized Lundquist number, $S_{g}=v_{A}\mu_{0}/k_{\perp}\eta_{g}$.
In the following discussion, we use two normalizations: the hat which
indicates normalization by Alfvén time/velocity and characteristic
field strengths ($\hat{\omega}=\omega\tau_{A}$, $\hat{L}=kL$, $\hat{v}=v/v_{A}\left(x=0\right)$,
$\hat{B}=B/B_{0}\left(x=0\right)$, $\hat{n}=n/n_{0}\left(x=0\right)$,
and $\hat{p}=p/v_{A}^{2}m_{i}n_{0}\left(x=0\right)$) and the overbar
which is a tearing specific normalization introduced in Sec.~\ref{sec:SysEqns}.

\section{Linearized Equations \label{sec:LinEqns}}

\noindent Following convention, we define $\hat{\xi}=\hat{\gamma}\hat{v}_{x}$
as the displacement vector and 
\begin{equation}
\hat{Q}=\hat{k}^{2}\hat{B}_{\parallel}-i\hat{k}_{\parallel}\hat{B_{x}^{\prime}}+i\hat{\lambda}_{0}\hat{B}_{x}\;,
\end{equation}
consistent with Ref.~\cite{Ahedo_09_PPaCF_55018} where $\lambda=\mu_{0}\mathbf{J}\cdot\mathbf{B}/B^{2}$
and $\hat{k}_{\parallel}=\mathbf{k}\cdot\mathbf{B}_{0}/kB_{0}$. After
linearization and applying the assumptions of Sec.~\ref{sec:ModelEqnsOrdering},
the radial induction equation becomes
\begin{equation}
\hat{\gamma}_{e}\hat{B}_{x}=i\hat{k}_{\parallel}\hat{\gamma}\hat{\xi}+\hat{k}_{\parallel}\hat{d}_{i}\hat{Q}+S_{g}^{-1}\hat{B_{x}^{\prime\prime}}\;.\label{eq:linRadInd}
\end{equation}
The left side of this equation is a term representing the rate-of-change
of $\hat{B}_{x}$. The notation $\hat{\gamma}_{i}=\hat{\gamma}+i\hat{\mathbf{k}}\cdot\mathbf{\hat{v}}_{0}$,
and $\hat{\gamma}_{e}=\hat{\gamma}+i\hat{\mathbf{k}}\cdot\mathbf{\hat{v}}_{e0}\simeq\hat{\gamma}_{i}-i\hat{\omega}_{*}$
gathers the advective and temporal-derivative contributions into a
single term. The terms on the right side of Eqn.~(\ref{eq:linRadInd})
result from the $\mathbf{v}\times\mathbf{B}$, Hall, and resistive/inertial
terms, respectively. Contributions from the $\nabla p_{e}$ term vanish.
Other than ignoring flow shear and applying our ordering to resistive/inertial
term, Eqn.~(\ref{eq:linRadInd}) is exact. 

The location where $\mathbf{k}\cdot\mathbf{B}_{0}=0$ is the resonant
magnetic-flux surface. Away from the resonant surface the contribution
from the $\mathbf{v}\times\mathbf{B}$ term dominates and all other
terms may be neglected. When fluid decoupling and/or drift effects
are significant, the Hall term dominates near the resonant surface.
At the resonant surface the $\mathbf{v}\times\mathbf{B}$ and Hall
terms vanish and thus the resistive and inertial contributions must
be retained. Our calculations assume the resonant surface is located
at $x=0$. The standard treatment of these equations is to apply a
boundary-layer analysis, where the ideal-MHD equations describe the
solution in the outer region (away from the resonant surface), and
the full model is used in the inner layer near the resonant surface.
These solutions are matched using the discontinuity in the logarithmic
derivative of the perturbed radial magnetic field of the outer solution
($\Delta^{\prime}$),
\begin{equation}
\Delta^{\prime}=\underset{\epsilon\rightarrow0}{lim}\frac{\left.\tilde{B}_{x}^{\prime}\left(x\right)\right|_{x=-\epsilon}^{x=\epsilon}}{\tilde{B}_{x}\left(0\right)}\;,\label{eq:deltaPrime}
\end{equation}
where the prime indicates a partial derivative with respect to $x$.
With a resistive-MHD model, an equilibrium is tearing unstable ($\gamma>0$)
if $\Delta^{\prime}>0$ \cite{Furth_63_PoF_459}; thus $\Delta^{\prime}$
is both a matching and stability parameter. We assume that $\Delta^{\prime}\delta\sim\mathcal{O}\left(1\right)$
and thus $\hat{B_{x}^{\prime}}\sim\hat{B}_{x}$, as follows from Eqn.
(\ref{eq:deltaPrime}). Expanding $\hat{B}_{x}$ at $x=0$, 
\begin{equation}
\hat{B}_{x}=\hat{B}_{x}\left(0\right)+\hat{B_{x}^{\prime}}\left(0\right)\hat{x}+...\;,
\end{equation}
and noting that $\hat{x}\sim\mathcal{O}\left(\epsilon\right)$ allows
us to treat $\hat{B}_{x}$ as a constant - an assumption known as
the constant-$\psi$ approximation. Derivatives of other perturbed
fields are assumed to raise the relative size of the field by $\epsilon^{-1}$,
e.g. $\epsilon^{2}\hat{\xi^{\prime\prime}}\sim\hat{\xi}$ and $\epsilon\hat{B_{x}^{\prime\prime}}\sim\hat{B}_{x}$.
This approximation results from the large, localized gradients of
perturbed fields within the boundary layer. Consider, for example,
that the reconnecting inflows of the tearing mode produce a displacement
vector that changes sign across the boundary layer.

After linearization, the parallel induction equation becomes
\begin{multline}
\hat{\gamma}_{e}\hat{B}_{\parallel}=-\hat{\nabla}_{\perp}\cdot\hat{\mathbf{v}}+\hat{\omega}_{*}\frac{\hat{\gamma}\hat{\xi}}{\hat{d}_{i}}-\left(i\hat{\omega}_{*}+i\hat{\omega}_{*n}\frac{\Gamma}{\hat{c}_{s}^{2}}\right)\hat{Q}+\hat{k}_{\parallel}\hat{d}_{i}\left[\hat{B_{x}^{\prime\prime}}+i\hat{k}_{\parallel}\hat{B_{\parallel}^{\prime}}-\hat{B}_{x}\right]\\
-i\hat{\omega}_{*i}\hat{n}-i\hat{\omega}_{*n}\frac{\Gamma}{\hat{c}_{s}^{2}}\hat{p}_{e}+S_{g}^{-1}\hat{B_{\parallel}^{\prime\prime}}\;.\label{eq:linParInd}
\end{multline}
where $\omega_{*\alpha}$ is a species diamagnetic-drift frequency
($kp_{\alpha0}^{\prime}/n_{0}eB_{0}$), $\omega_{*}$ is the total
diamagnetic-drift frequency ($\omega_{*i}+\omega_{*e}$), $\omega_{*n}$
is the density-gradient drift $\left(kT_{0}n_{0}^{\prime}/n_{0}eB_{0}\right)$
and $\nabla_{\perp}=\nabla-ik_{\parallel}\hat{\mathbf{b}}$. The first
two pairs of terms on the right side are the contributions from the
$\mathbf{v}\times\mathbf{B}$ and Hall terms, respectively. The terms
involving $\hat{n}$ and $\hat{p}_{e}$ result from the $\nabla p_{e}$
term and the last term is the effect of resistivity and electron inertia. 

The components of the linearized momentum equation are 
\begin{equation}
\hat{\gamma}_{i}\hat{\gamma}\hat{\xi}=\frac{\hat{\omega}_{*}}{\hat{d}_{i}}\hat{B}_{\parallel}+i\hat{k}_{\parallel}\hat{B}_{x}-\hat{B_{\parallel}^{\prime}}-\hat{p^{\prime}}-\left(\hat{\nabla}\cdot\hat{\mathbf{\Pi}}\right)_{x}\;,\label{eq:linRadMom}
\end{equation}
\begin{equation}
\hat{\gamma}_{i}\hat{v}_{\perp}=-i\hat{Q}-i\hat{p}-\left(\hat{\nabla}\cdot\hat{\mathbf{\Pi}}\right)_{\perp}\;,\label{eq:linPerpMom}
\end{equation}
and
\begin{equation}
\hat{\gamma}_{i}\hat{v}_{\parallel}=-\frac{\hat{\omega}_{*}}{\hat{d}_{i}}\hat{B}_{x}-i\hat{k}_{\parallel}\hat{p}-\left(\hat{\nabla}\cdot\hat{\mathbf{\Pi}}\right)_{\parallel}\;.\label{eq:linParMom}
\end{equation}
The perpendicular and parallel components (Eqns.~(\ref{eq:linPerpMom})
and (\ref{eq:linParMom})) are used to construct an expression for
$\hat{\nabla}\cdot\hat{\mathbf{v}}$. The first terms on the right
side of Eqns.~(\ref{eq:linRadMom}) and (\ref{eq:linParMom}) are
drift contributions from $\mathbf{J}\times\mathbf{B}$.

The linearized continuity, ion-energy and electron-energy equations
are
\begin{equation}
\hat{\gamma}_{i}\hat{n}=-\hat{\omega}_{*n}\frac{\Gamma}{\hat{c}_{s}^{2}}\frac{\hat{\gamma}\hat{\xi}}{\hat{d}_{i}}-\hat{\nabla}\cdot\hat{\mathbf{v}}\;,\label{eq:linCont}
\end{equation}
\begin{equation}
\hat{\gamma}_{i}\hat{p}_{i}=-\hat{\omega}_{*i}\frac{\hat{\gamma}\hat{\xi}}{\hat{d}_{i}}-\hat{c}_{si}^{2}\hat{\nabla}\cdot\hat{\mathbf{v}}-\left(\Gamma-1\right)\hat{\nabla}\cdot\hat{\mathbf{q}}_{i}\;,\label{eq:preLinIonEnergy}
\end{equation}
and
\begin{multline}
\hat{\gamma}_{pe}\hat{p}_{e}=-\hat{\omega}_{*e}\frac{\hat{\gamma}\hat{\xi}}{\hat{d}_{i}}-\hat{c}_{se}^{2}\hat{\nabla}\cdot\hat{\mathbf{v}}+\sigma_{pe}\left(i\omega_{*e}-\Gamma f_{Te}i\omega_{*n}\right)\left(\hat{Q}-i\hat{\lambda}_{0}\hat{B}_{x}\right)\\
-\sigma_{pe}\hat{c}_{se}^{2}\left(i\hat{\omega}_{*}+i\hat{k}_{\parallel}\hat{\lambda}_{0}\hat{d}_{i}\right)\hat{n}-\left(\Gamma-1\right)\hat{\nabla}\cdot\hat{\mathbf{q}}_{e}\;,\label{eq:preLinElecEnergy}
\end{multline}
respectively. Advection by fast, parallel, electron flows can be computationally
expensive to model in extended-MHD computations. A common computational
practice is to use the bulk flow in the advective term of the electron-energy
equation which circumvents the large computational cost of the fast
electron flows. To allow for a systematic study of the effect of different
advective models, we introduce the $\sigma_{pe}$ and $\hat{\gamma}_{pe}$
notation. If the advective term uses the bulk flow then $\hat{\gamma}_{pe}=\hat{\gamma}_{i}$
and $\sigma_{pe}=0$, whereas advection by the electron flow leads
to $\hat{\gamma}_{pe}=\hat{\gamma}_{e}$ and $\sigma_{pe}=1$. To
compute the linearized cross heat-flux contributions we first expand
the heat-flux vector as
\begin{multline}
\nabla\cdot\mathbf{q}_{\alpha}=\nabla\cdot\left[\frac{5p_{\alpha}}{2q_{\alpha}B}\hat{\mathbf{b}}\times\nabla T_{\alpha}\right]=\frac{5p_{\alpha}}{2nq_{\alpha}B^{2}}\left[\mu_{0}\mathbf{J}\cdot\left(p_{\alpha}\frac{\nabla n}{n}-\nabla p_{\alpha}\right)-2\lambda\mathbf{B}\cdot\left(p_{\alpha}\frac{\nabla n}{n}-\nabla p_{\alpha}\right)\right]\\
+\frac{5p_{\alpha}}{nq_{\alpha}B^{4}}\left[\frac{p_{\alpha}}{n}\left(\mathbf{B}\times\nabla n\right)-\mathbf{B}\times\nabla p_{\alpha}\right]\cdot\mathbf{B}\cdot\nabla\mathbf{B}+\frac{5p_{\alpha}}{2n^{2}q_{\alpha}B^{2}}\mathbf{B}\cdot\left(\nabla p_{\alpha}\times\nabla n\right)\;.
\end{multline}
 Noting that $\mathbf{J}_{0}\cdot\nabla f_{0}$, $\mathbf{B}_{0}\cdot\nabla f_{0}$,
$\mathbf{B}_{0}\cdot\nabla\mathbf{B}_{0}$, and $\nabla f_{0}\times\nabla g_{0}$
vanish for our slab configuration, we may assume the coefficients
of these terms are equilibrium quantities during linearization. After
linearization and ordering (specifically, we drop terms where $\hat{\omega}_{*}>>\hat{k}_{\parallel}\hat{\lambda}_{0}\hat{d}_{i}$),
we find
\begin{equation}
\left(\Gamma-1\right)\hat{\nabla}\cdot\hat{\mathbf{q}}_{\alpha}=i\hat{\omega}_{*q1\alpha}\frac{\hat{c}_{s\alpha}^{2}}{\Gamma}\hat{n}-i\hat{\omega}_{*q\alpha}\hat{p}_{\alpha}-\left(\hat{\gamma}_{\alpha}-i\hat{\omega}_{*q\alpha}\right)\hat{c}_{s\alpha}^{2}C_{q\alpha}\left(\hat{Q}-i\hat{\lambda}_{0}\hat{B}_{x}+2i\hat{k}_{\parallel}\hat{B_{x}^{\prime}}\right)\;,\label{eq:linHeatFlux}
\end{equation}
where
\begin{equation}
C_{q\alpha}=\sigma_{q\alpha}\frac{i\hat{\omega}_{*\alpha}-f_{T\alpha}i\hat{\omega}_{*n}}{\hat{\gamma}_{\alpha}-i\hat{\omega}_{*q\alpha}}\;,
\end{equation}
\begin{equation}
i\hat{\omega}_{*q1\alpha}=\sigma_{q\alpha}\left(\Gamma i\hat{\omega}_{*\alpha}+\hat{c}_{s\alpha}^{2}i\hat{\omega}_{*}\right)\;,
\end{equation}
and
\begin{equation}
i\hat{\omega}_{*q\alpha}=\sigma_{q\alpha}f_{T\alpha}\left(\Gamma i\hat{\omega}_{*n}+\hat{c}_{s}^{2}i\hat{\omega}_{*}\right)\;.
\end{equation}
Again we introduce $\sigma_{q\alpha}$ as a marker with value $\sigma_{qi}=-\sigma_{qe}=\left(5/2\right)\left(\Gamma-1\right)/\Gamma$
when the cross heat flux is included in the model and $\sigma_{q\alpha}=0$
when it is not. Eqns.~(\ref{eq:preLinIonEnergy}), (\ref{eq:preLinElecEnergy})
and (\ref{eq:linHeatFlux}) may be combined to produce expressions
for $\hat{p}=\hat{p}_{i}+\hat{p}_{e}$ and $\hat{p}_{e}$. Thus
\begin{equation}
\hat{p}=-E_{t}\frac{\hat{\gamma}\hat{\xi}}{\hat{d}_{i}}-\frac{\hat{c}_{sp}^{2}}{\hat{\gamma}_{i}}\hat{\nabla}\cdot\hat{\mathbf{v}}+\left(C_{pe}+\hat{c}_{sq}^{2}\right)\hat{Q}-\left(C_{pe}+\hat{c}_{sq}^{2}\right)i\hat{\lambda}_{0}\hat{B}_{x}+2\hat{c}_{sq}^{2}i\hat{k}_{\parallel}\hat{B_{x}^{\prime}}\;,\label{eq:linEnergy}
\end{equation}
and
\begin{equation}
\hat{p}_{e}=-E_{e}\frac{\hat{\gamma}\hat{\xi}}{\hat{d}_{i}}-\frac{\hat{c}_{spe}^{2}}{\hat{\gamma}_{i}}\hat{\nabla}\cdot\hat{\mathbf{v}}+\left(C_{pe}+\hat{c}_{sqe}^{2}\right)\hat{Q}-\left(C_{pe}+\hat{c}_{sqe}^{2}\right)i\hat{\lambda}_{0}\hat{B}_{x}+2\hat{c}_{sqe}^{2}i\hat{k}_{\parallel}\hat{B_{x}^{\prime}}\;,\label{eq:linElecEnergy}
\end{equation}
where $\hat{c}_{sqe}^{2}=C_{qe}\hat{c}_{se}^{2}$, $\hat{c}_{sq}^{2}=C_{qe}\hat{c}_{se}^{2}+C_{qi}\hat{c}_{si}^{2}$,
$\hat{c}_{sp}^{2}=\hat{c}_{spe}^{2}+\hat{c}_{spi}^{2}$,
\begin{equation}
\hat{c}_{spi}^{2}=\hat{c}_{si}^{2}\frac{\hat{\gamma}_{i}-i\hat{\omega}_{*q1i}/\Gamma}{\hat{\gamma}_{i}-i\hat{\omega}_{*qi}}\;,
\end{equation}
\begin{equation}
\hat{c}_{spe}^{2}=\hat{c}_{se}^{2}\frac{\hat{\gamma}_{pe}-i\hat{\omega}_{*q1e}/\Gamma}{\hat{\gamma}_{pe}-i\hat{\omega}_{*qe}}\,,
\end{equation}
\begin{equation}
C_{pe}=\sigma_{pe}\frac{i\hat{\omega}_{*e}-\Gamma f_{Te}i\hat{\omega}_{*n}}{\hat{\gamma}_{pe}-i\hat{\omega}_{*qe}}\;,
\end{equation}
\begin{equation}
E_{i}=\left(\hat{\gamma}_{i}-i\hat{\omega}_{*qi}\right)^{-1}\left(\hat{\omega}_{*i}-f_{Ti}\frac{i\hat{\omega}_{*n}\hat{\omega}_{*q1i}}{\hat{\gamma}_{i}}\right)\;,
\end{equation}
\begin{equation}
E_{e}=\left(\hat{\gamma}_{pe}-i\hat{\omega}_{*qe}\right)^{-1}\left(\hat{\omega}_{*e}-f_{Te}\frac{i\hat{\omega}_{*n}\hat{\omega}_{*q1e}}{\hat{\gamma}_{i}}-\frac{\hat{\omega}_{*n}f_{Te}\Gamma\sigma_{pe}i\hat{\omega}_{*}}{\hat{\gamma}_{i}}\right)\;,
\end{equation}
and $E_{t}=E_{i}+E_{e}$.

\section{System of Equations \label{sec:SysEqns}}

\noindent We next algebraically reduce Eqns.~(\ref{eq:linRadInd}),
(\ref{eq:linParInd})-(\ref{eq:linCont}), (\ref{eq:linEnergy}) and
(\ref{eq:linElecEnergy}) from a system of eight equations to a system
of five. These five equations use $\hat{B}_{x}$, $\hat{\nabla}\cdot\hat{\mathbf{v}}$,
$\hat{Q}$, $\hat{\xi}$, and $\hat{v}_{\parallel}$ as primary variables.
Two of these are unmodified from the system of eight: the radial induction
equation, Eqn.~(\ref{eq:linRadInd}), and the parallel velocity equation,
Eqn.~(\ref{eq:linParMom}). One is slightly modified: the parallel
induction equation provides an expression for $\hat{Q}$ after $\hat{n}$
and $\hat{p}$ are eliminated. And two new equations are derived:
an expression for $\hat{\nabla}\cdot\hat{\mathbf{v}}$ and a parallel
vorticity equation which governs $\hat{\xi}$.

Eqns.~(\ref{eq:linPerpMom}) and (\ref{eq:linParMom}) are combined
to provide an expression for $\hat{\nabla}\cdot\hat{\mathbf{v}}$,
\begin{equation}
\hat{p}=\hat{\gamma}_{i}\hat{\nabla}\cdot\hat{\mathbf{v}}-\hat{\gamma}_{i}\hat{\gamma}\hat{\xi^{\prime}}-\hat{Q}+\frac{i\hat{k}_{\parallel}\hat{\omega}_{*}}{\hat{d}_{i}}\hat{B}_{r}+i\left(\hat{\nabla}\cdot\hat{\mathbf{\Pi}}_{gv}\right)_{\perp}+i\hat{k}_{\parallel}\left(\hat{\nabla}\cdot\hat{\mathbf{\Pi}}_{gv}\right)_{\parallel}\;.\label{eq:pDivV}
\end{equation}
After multiplying by $\hat{\gamma}_{i}$ and substituting Eqn.~(\ref{eq:linEnergy})
for $\hat{p}$, 
\begin{multline}
\hat{c}_{sp}^{2}\hat{\nabla}\cdot\hat{\mathbf{v}}=\hat{\gamma}_{i}^{2}\hat{\gamma}\hat{\xi^{\prime}}-\hat{\gamma}_{i}E_{t}\frac{\hat{\gamma}\hat{\xi}}{\hat{d}_{i}}+\hat{\gamma}_{i}\left(1+C_{pe}+\hat{c}_{sq}^{2}\right)\hat{Q}-\hat{\gamma}_{i}\left(\frac{\hat{k}_{\parallel}\hat{\omega}_{*}}{\hat{d}_{i}\hat{\lambda}_{0}}+C_{pe}+\hat{c}_{sq}^{2}\right)i\hat{\lambda}_{0}\hat{B}_{x}+2\hat{\gamma}_{i}\hat{c}_{sq}^{2}i\hat{k}_{\parallel}\hat{B_{x}^{\prime}}\\
-i\hat{\gamma}_{i}\left[\left(\hat{\nabla}\cdot\hat{\mathbf{\Pi}}_{gv}\right)_{\perp}+\hat{k}_{\parallel}\left(\hat{\nabla}\cdot\hat{\mathbf{\Pi}}_{gv}\right)_{\parallel}\right]\;.\label{eq:s5DivV}
\end{multline}
The inertial contributions ($\hat{\gamma}_{i}^{2}\hat{\nabla}\cdot\hat{\mathbf{v}}$)
are dropped as they are small compared to the $\hat{c}_{sp}^{2}\hat{\nabla}\cdot\hat{\mathbf{v}}$
term from $\hat{p}$ in Eqn.~(\ref{eq:linEnergy}). Without drift
and FLR effects only the first and third terms on the right side contribute
to $\hat{\nabla}\cdot\hat{\mathbf{v}}$. The second term on the right
side is a drift-like term from $\tilde{\mathbf{v}}\cdot\nabla p$
and $\tilde{\mathbf{v}}\cdot\nabla n$ and the remaining terms are
contributions from electron advection ($\sim C_{pe}$), cross heat
flux ($\sim c_{sq}^{2}$) and ion gyroviscosity. 

After eliminating $\hat{B}_{\parallel}$, $\hat{n}$ and $\hat{p}$
from the parallel induction equation, Eqn.~(\ref{eq:linParInd}),
we find
\begin{multline}
\left(\hat{\gamma}_{i}-i\hat{\omega}_{*}\right)\hat{Q}=\left(A-1\right)\hat{\nabla}\cdot\hat{\mathbf{v}}+i\hat{k}_{\parallel}\hat{v}_{\parallel}+\hat{k}_{\parallel}\hat{d}_{i}\hat{B_{x}^{\prime\prime}}+\left(\hat{\omega}_{*}+i\hat{\omega}_{*n}\frac{\Gamma}{\hat{c}_{s}^{2}}E_{n}\right)\frac{\hat{\gamma}\hat{\xi}}{\hat{d}_{i}}\\
-\left[i\hat{\omega}_{*}+i\hat{\omega}_{*n}\frac{\Gamma}{\hat{c}_{s}^{2}}\left(1+C_{pe}+\hat{c}_{sqe}^{2}\right)\right]\hat{Q}+i\hat{\omega}_{*n}\frac{\Gamma}{\hat{c}_{s}^{2}}\left(C_{pe}+\hat{c}_{sqe}^{2}\right)i\hat{\lambda}_{0}\hat{B}_{x}+S_{g}^{-1}\hat{Q^{\prime\prime}}\;,\label{eq:s5ParInd}
\end{multline}
where
\begin{equation}
A=\frac{i\hat{\omega}_{*i}}{\hat{\gamma}_{i}}+\Gamma\frac{\hat{c}_{spe}^{2}}{\hat{c}_{s}^{2}}\frac{i\hat{\omega}_{*n}}{\hat{\gamma}_{i}}\;,
\end{equation}
and
\begin{equation}
E_{n}=E_{e}+\frac{\hat{\omega}_{*i}}{\hat{\gamma}_{i}}\;.
\end{equation}
Without drift effects, all contributions from $\nabla p_{e}$ and
$\nabla n$ vanish (the latter of these results from the $1/ne$ factors
in Ohm's law). In particular, these contributions lead to the $A$,
$E_{n}$, $C_{pe}$ and $\hat{c}_{sq}^{2}$ factors in Eqn.~(\ref{eq:s5ParInd}).

The only unused equation from our original system of eight is the
radial momentum equation, Eqn.~(\ref{eq:linRadMom}). To find an
expression for $\hat{p}^{\prime}$ we take the derivative of Eqn.~(\ref{eq:pDivV}):
\begin{multline}
\hat{p}^{\prime}=-\hat{\gamma}_{i}\hat{\gamma}\hat{\xi^{\prime\prime}}-\hat{\omega}_{*n}\frac{\Gamma}{\hat{c}_{s}^{2}}\hat{\gamma}_{i}\frac{\hat{\gamma}\hat{\xi^{\prime}}}{\hat{d}_{i}}+\frac{\hat{\omega}_{*}}{\hat{d}_{i}}\hat{Q}-\hat{Q^{\prime}}+\frac{i\hat{k}_{\parallel}\hat{\omega}_{*}}{\hat{d}_{i}}\hat{B_{x}^{\prime}}+\left(\frac{\hat{\omega}_{*}}{\hat{d}_{i}}+\frac{\hat{\omega}_{*}^{2}\hat{k}_{\parallel}}{\hat{\lambda}_{0}\hat{d}_{i}^{2}}\right)i\hat{\lambda}_{0}\hat{B}_{x}\\
+i\left(\hat{\nabla}\cdot\hat{\mathbf{\Pi}}_{gv}\right)_{\perp}^{\prime}+i\hat{k}_{\parallel}\left(\hat{\nabla}\cdot\hat{\mathbf{\Pi}}_{gv}\right)_{\parallel}^{\prime}+i\hat{\lambda}_{0}\left(\hat{\nabla}\cdot\hat{\mathbf{\Pi}}_{gv}\right)_{\parallel}\;.
\end{multline}
Again, we ignore the inertial term ($\hat{\gamma}_{i}^{2}\hat{\nabla}\cdot\hat{\mathbf{v}}$).
Substituting into Eqn.~(\ref{eq:linRadMom}) and applying the tearing
ordering,
\begin{multline}
\hat{\gamma}_{i}\hat{\gamma}\hat{\xi^{\prime\prime}}=2\hat{k}_{\parallel}\hat{\lambda}_{0}\hat{Q}-i\hat{k}_{\parallel}\hat{B_{x}^{\prime\prime}}+\hat{\omega}_{*n}\frac{\Gamma}{\hat{c}_{s}^{2}}\hat{\gamma}_{i}\frac{\hat{\gamma}\hat{\xi^{\prime}}}{\hat{d}_{i}}-2\frac{i\hat{\omega}_{*}}{\hat{d}_{i}}\hat{\lambda}_{0}\hat{B}_{x}\\
-\left(\hat{\nabla}\cdot\hat{\mathbf{\Pi}}\right)_{r}-i\left(\hat{\nabla}\cdot\hat{\mathbf{\Pi}}_{gv}\right)_{\perp}^{\prime}-i\hat{k}_{\parallel}\left(\hat{\nabla}\cdot\hat{\mathbf{\Pi}}_{gv}\right)_{\parallel}^{\prime}-i\hat{\lambda}_{0}\left(\hat{\nabla}\cdot\hat{\mathbf{\Pi}}_{gv}\right)_{\parallel}\;.\label{eq:linParVort}
\end{multline}
Without drift and FLR effects, this equation becomes the standard
form of the parallel vorticity equation, $\hat{\gamma}_{i}\hat{\gamma}\hat{\xi^{\prime\prime}}\simeq-i\hat{k}_{\parallel}\hat{B_{x}^{\prime\prime}}$. 

We now have a system of five equations: Eqns.~(\ref{eq:linRadInd}),
(\ref{eq:linParMom}), (\ref{eq:s5DivV}), (\ref{eq:s5ParInd}), and
(\ref{eq:linParVort}). The discussion of the tearing-ordered contributions
from ion gyroviscosity are deferred until the next section. Without
these contributions, compressibility and parallel flows only couple
to this system through the parallel induction equation, Eqn.~(\ref{eq:s5ParInd}).
Thus in the single-fluid regime where the Hall effect and ion gyroviscosity
may be ignored, only two equations, the radial induction and parallel
vorticity equations, are required for a solution.

\subsection{Considerations of Ion Gyroviscosity\label{sub:IonGV}}

With tearing-ordered gyroviscous contributions, the compressibility
equation (Eqn.~(\ref{eq:s5DivV})) becomes 
\begin{multline}
\hat{c}_{sp}^{2}\hat{\nabla}\cdot\hat{\mathbf{v}}=\hat{\gamma}_{i}E_{t}\frac{\hat{\gamma}\hat{\xi}}{\hat{d}_{i}}+\hat{\gamma}_{i}\left(1+C_{pe}+\hat{c}_{sq}^{2}\right)\hat{Q}-\hat{\gamma}_{i}\left(\frac{\hat{k}_{\parallel}\hat{\omega}_{*}}{\hat{d}_{i}\hat{\lambda}_{0}}+C_{pe}+\hat{c}_{sq}^{2}\right)i\hat{\lambda}_{0}\hat{B}_{x}+2\hat{\gamma}_{i}\hat{c}_{sq}^{2}i\hat{k}_{\parallel}\hat{B_{x}^{\prime}}\\
+\hat{\gamma}_{i}\hat{\gamma}_{gvi}\hat{\gamma}\hat{\xi^{\prime}}-\sigma_{gv}\frac{i\hat{\gamma}_{i}}{2}\frac{\hat{c}_{si}^{2}}{\Gamma}\hat{d}_{i}\hat{\gamma}\hat{\xi^{\prime\prime}}\;,
\end{multline}
and the parallel-momentum equation (Eqn.~(\ref{eq:linParMom})) becomes
\begin{multline}
\hat{\gamma}_{gvi}\hat{v}_{\parallel}=-\frac{\hat{\omega}_{*}}{\hat{d}_{i}}\hat{B}_{x}+\frac{i\hat{k}_{\parallel}\hat{c}_{sp}^{2}}{\hat{\gamma}_{i}}\hat{\nabla}\cdot\hat{\mathbf{v}}-i\hat{k}_{\parallel}\hat{\gamma}_{i}\left(C_{pe}+\hat{c}_{sq}^{2}\right)\hat{Q}+i\hat{k}_{\parallel}E_{t}\frac{\hat{\gamma}\hat{\xi}}{\hat{d}_{i}}\\
-\sigma_{gv}\frac{\hat{c}_{si}^{2}}{\Gamma}\hat{\lambda}_{0}\hat{d}_{i}\hat{\gamma}\hat{\xi^{\prime}}-\sigma_{gv}\frac{\hat{c}_{si}^{2}}{\Gamma}\hat{d}_{i}\hat{k}_{\parallel}\hat{\gamma}\hat{\xi^{\prime\prime}}\;,\label{eq:linParMomGV}
\end{multline}
where $\sigma_{gv}$ is a marker for ion gyroviscosity (set to unity
when gyroviscosity is included and otherwise zero), the modified ion
gyroviscous frequency is 
\begin{equation}
\hat{\gamma}_{gvi}=\hat{\gamma}_{ExB}+i\hat{\omega}_{*i}-\sigma_{gv}\left(i\hat{\omega}_{*i}-i\hat{\omega}_{*}\frac{\hat{c}_{si}^{2}}{\Gamma}\right)\;,
\end{equation}
and $\hat{\gamma}_{ExB}$ is the doppler-shifted growth rate. The
tearing-ordered ion-gyroviscous contributions to parallel-vorticity
equation (Eqn.~(\ref{eq:linParVort})) are 
\begin{multline}
-\left(\hat{\nabla}\cdot\hat{\mathbf{\Pi}}\right)_{r}-i\left(\hat{\nabla}\cdot\hat{\mathbf{\Pi}}_{gv}\right)_{\perp}^{\prime}-i\hat{k}_{\parallel}\left(\hat{\nabla}\cdot\hat{\mathbf{\Pi}}_{gv}\right)_{\parallel}^{\prime}-i\hat{\lambda}_{0}\left(\hat{\nabla}\cdot\hat{\mathbf{\Pi}}_{gv}\right)_{\parallel}=\frac{i\hat{\omega}_{*}\hat{\omega}_{*i}}{\hat{d}_{i}}\hat{\nabla}\cdot\hat{\mathbf{v}}\\
-2i\hat{\omega}_{*}\left(\hat{\omega}_{*i}+\hat{\omega}_{*}\frac{\hat{c}_{si}^{2}}{\Gamma}\right)\frac{\hat{\gamma}\hat{\xi^{\prime}}}{\hat{d}_{i}}-i\frac{\hat{c}_{si}^{2}}{\Gamma}\hat{d}_{i}\left(\left(\hat{\nabla}\cdot\hat{\mathbf{v}}\right)^{\prime\prime}+i\hat{k}_{\parallel}\hat{v_{\parallel}^{\prime\prime}}\right)-\left(i\hat{\omega}_{*i}+i\hat{\omega}_{*}\frac{\hat{c}_{si}^{2}}{\Gamma}\right)\hat{\gamma}\hat{\xi^{\prime\prime}}\;.\label{eq:pvGV}
\end{multline}
The $i\hat{\omega}_{*i}\hat{\gamma}\hat{\xi^{\prime\prime}}$ term
produces the standard gyroviscous cancellation and cancels the advective
diamagnetic drift, however, as there are many additional terms in
the this equation, this cancellation is inexact. The $i\hat{\omega}_{*}\hat{c}_{si}^{2}\hat{\gamma}\hat{\xi^{\prime\prime}}/\Gamma$
term is the result of a drift proportional to the gradient of the
magnetic field as previously discussed in detail for tearing in a
cylindrical pinch configuration \cite{King_11_PoP_42303} (it has
been re-characterized in terms of $\omega_{*}$ through equilibrium
force balance). Combining Eqns.~(\ref{eq:linParVort}) and (\ref{eq:pvGV})
and again applying the tearing ordering gives
\begin{equation}
-\hat{\gamma}_{gvi}\hat{\gamma}\hat{\xi^{\prime\prime}}=2\hat{k}_{\parallel}\hat{\lambda}_{0}\hat{Q}-i\hat{k}_{\parallel}\hat{B_{x}^{\prime\prime}}-2\frac{i\hat{\omega}_{*}}{\hat{d}_{i}}\hat{\lambda}_{0}\hat{B}_{x}-i\sigma_{gv}\frac{\hat{c}_{si}^{2}}{\Gamma}\hat{d}_{i}\left(\left(\hat{\nabla}\cdot\hat{\mathbf{v}}\right)^{\prime\prime}+i\hat{k}_{\parallel}\hat{v_{\parallel}^{\prime\prime}}\right)\;.\label{eq:parParVortGV}
\end{equation}
The last two terms on the right side of Eqn.~(\ref{eq:parParVortGV})
raise the differential order of the system of equations. Without these
contributions, compressibility and parallel flows can be eliminated
algebraically from the parallel induction equation, Eqn.~(\ref{eq:s5ParInd}),
which is the only other location where these variables enter the system
of equations. We do not presently have a solution to the system of
equations with ion gyroviscosity, and thus we proceed without the
full contributions. 

Prior work typically includes only the standard gyroviscous cancellation
as a model of ion gyroviscosity. Although we can not justify this
approximation from a tearing-ordered-equations stand point, we retain
the $\hat{\gamma}_{gvi}$ terms as is in order to facilitate comparison.
The two relevant limits are then without gyroviscosity ($\hat{\gamma}_{gvi}\rightarrow\hat{\gamma}_{i}$),
and with the exact gyroviscous cancellation ($\hat{\gamma}_{gvi}\rightarrow\hat{\gamma}_{ExB}$).

\subsection{Tearing Normalized System of Equations}

Without ion gyroviscosity, compressibility and parallel flow can be
eliminated algebraically. Substituting Eqns.~(\ref{eq:linParMom})
and (\ref{eq:s5DivV}) into Eqn.~(\ref{eq:s5ParInd}) we find
\begin{equation}
\hat{\tau}_{Q}\hat{Q}=\hat{k}_{\parallel}\hat{d}_{i}\hat{B_{x}^{\prime\prime}}+S_{g}^{-1}\hat{Q^{\prime\prime}}-\frac{\hat{k}_{\parallel}^{2}}{\hat{\gamma}_{gvi}}\hat{Q}+\left(\hat{\tau}_{B}-\frac{\hat{k}_{\parallel}\hat{\omega}_{*}}{\hat{\gamma}_{gvi}\hat{d}_{i}\hat{\lambda}_{0}}\right)i\hat{\lambda}_{0}\hat{B}_{x}+\hat{\tau}_{\xi}\frac{\hat{\gamma}\hat{\xi}}{\hat{d}_{i}}\;,\label{eq:s3ParInd}
\end{equation}
where
\begin{equation}
\hat{\tau}_{Q}=\hat{\gamma}_{i}+i\hat{\omega}_{*n}\frac{\Gamma}{\hat{c}_{s}^{2}}\left(1+C_{pe}+\hat{c}_{sqe}^{2}\right)-\frac{\hat{\gamma}_{i}}{\hat{c}_{sp}^{2}}\left(1+C_{pe}+\hat{c}_{sq}^{2}\right)\left(A-1\right)\;,
\end{equation}
\begin{equation}
\hat{\tau}_{B}=i\hat{\omega}_{*n}\frac{\Gamma}{\hat{c}_{s}^{2}}\left(C_{pe}+\hat{c}_{sqe}^{2}\right)+\hat{\gamma}_{i}\left(C_{pe}+\hat{c}_{sq}^{2}\right)\frac{\left(A-1\right)}{\hat{c}_{sp}^{2}}\;,
\end{equation}
and
\begin{equation}
\hat{\tau}_{\xi}=\hat{\omega}_{*}+i\hat{\omega}_{*n}\frac{\Gamma}{\hat{c}_{s}^{2}}E_{n}-\frac{\left(A-1\right)}{\hat{c}_{sp}^{2}}\hat{\gamma}_{i}E_{t}\;.
\end{equation}
Equations (\ref{eq:linRadInd}), (\ref{eq:parParVortGV}) (with $\sigma_{gv}=0$),
and (\ref{eq:s3ParInd}) now comprise our system of equations for
$\hat{B}_{x}$, $\hat{Q}$ and $\hat{\xi}$. The first two terms on
the right side of Eqn.~(\ref{eq:s3ParInd}) are the contributions
from the Hall term and resistive diffusion, respectively; the remaining
terms result from a combination of compressibility, parallel flows,
$\nabla p_{e}$ contributions, inertia and the $\mathbf{v}\times\mathbf{B}$
term. Compressibility and parallel flows contribute the $\hat{k}_{\parallel}^{2}$
and $\hat{\omega}_{*}$ terms on the right side of Eqn.~(\ref{eq:s3ParInd})
as well as the $\hat{\gamma}_{i}/\hat{c}_{sp}^{2}$ terms in the $\bar{\tau}$
factors. The $\nabla p_{e}$ term in Ohm's law contributes the $\hat{\omega}_{*n}/\hat{c}_{s}^{2}$
terms in the $\bar{\tau}$ factors. 

With the constant-$\psi$ approximation, where $\hat{B}_{x}$ is assumed
constant within the small tearing layer, Eqn.~(\ref{eq:parParVortGV})
is used to eliminate $\hat{B_{x}^{\prime\prime}}$; which results
in a system of two coupled equations for $\hat{Q}$ and $\hat{\xi}$.
We use a tearing normalization for these equations similar to Ref.~\cite{Ahedo_09_PPaCF_55018}
with the dimensionless variables,
\begin{equation}
\bar{x}=\frac{\hat{x}}{\hat{d}_{0}}\;,\quad\bar{\xi}=\frac{i\hat{k_{\parallel}^{\prime}}\hat{d}_{0}\hat{\gamma}\hat{\xi}}{\hat{B}_{r}\left(0\right)\hat{\gamma}_{e}}\;,\quad\bar{Q}=\frac{\hat{k_{\parallel}^{\prime}}\hat{d}_{0}\hat{d}_{i}\hat{Q}}{\hat{B}_{r}\left(0\right)\hat{\gamma}_{e}}\;,
\end{equation}
and the dimensionless parameters,
\begin{equation}
\hat{d}_{0}=\left(\frac{\hat{\gamma}_{ExB}}{\hat{k_{\parallel}^{\prime}}^{2}S_{g}}\right)^{1/4}\;,\quad\bar{\sigma}^{2}=\frac{\hat{\gamma}_{ExB}^{2}\hat{d}_{i}^{2}}{\hat{k_{\parallel}^{\prime}}^{2}\hat{d}_{0}^{4}}=\hat{\gamma}_{ExB}\hat{d}_{i}^{2}S_{g}\;,\quad\bar{R}=\frac{\hat{\gamma}_{gvi}}{\hat{\gamma}_{ExB}}\;,\quad\bar{\Lambda}=\frac{i\hat{\omega}_{*}}{\hat{\gamma}_{e}}\frac{\hat{\gamma}_{ExB}}{\hat{\gamma}_{gvi}}\;,
\end{equation}
\begin{equation}
\bar{\tau}_{Q}=\frac{\hat{\gamma}_{ExB}}{\left(\hat{k_{\parallel}^{\prime}}\hat{d}_{0}\right)^{2}}\hat{\tau}_{Q}\;,\quad\bar{\tau}_{\xi}=\frac{i\hat{\gamma}_{ExB}}{\left(\hat{k_{\parallel}^{\prime}}\hat{d}_{0}\right)^{2}}\hat{\tau}_{\xi}\;,\; and\quad\bar{\tau}_{B}=\frac{i\hat{\gamma}_{ExB}\hat{d}_{i}}{\hat{\gamma}_{e}\hat{d}_{0}}\left(\hat{\tau}_{B}+2i\hat{\omega}_{*}\right)\;.
\end{equation}
With this normalization, $\bar{\sigma}$ is the ion skin depth, $d_{i}$,
normalized to the tearing skin depth, $\delta=\left(S_{g}\hat{\gamma}\right)^{-1/2}$.
Validity of a first-order FLR model requires $\rho_{i}<\delta$. A
good rule of thumb for plasmas with comparable ion and electron temperatures
is to use the ion sound gyroradius, $\rho_{s}=c_{s}/\omega_{ci}$,
and require $\rho_{s}/\delta=\hat{c}_{s}\bar{\sigma}=\sqrt{\beta}\bar{\sigma}<1$.
After expanding $k_{\parallel}$ and retaining only the leading order
term in $x$, $k_{\parallel}^{\prime}x$, 
\begin{equation}
\bar{R}\frac{\partial^{2}\bar{\xi}}{\partial\bar{x}^{2}}=\bar{x}^{2}\left(\bar{\xi}+\bar{Q}\right)-\bar{x}\;,\label{eq:NormIon}
\end{equation}
and
\begin{equation}
\frac{\partial^{2}\bar{Q}}{\partial\bar{x}^{2}}=\left(\bar{R}^{-1}\bar{x}^{2}+\bar{\tau}_{Q}\right)\bar{Q}+\bar{R}\bar{\sigma}^{2}\frac{\partial^{2}\bar{\xi}}{\partial\bar{x}^{2}}+\bar{\tau}_{\xi}\bar{\xi}-\bar{\tau}_{B}+\bar{\Lambda}\bar{x}\label{eq:NormElec}
\end{equation}
compose the system of second-order coupled equations. 

Equation (\ref{eq:NormIon}), a combination of the radial-induction
and parallel-vorticity equations, governs the ion dynamics and is
composed of the contribution from resistivity on the left side, and
the contributions from the $\mathbf{v}\times\mathbf{B}$, Hall , and
inertial terms, respectively, on the right side. In the single-fluid
limit where the Hall term ($\bar{x}^{2}\bar{Q}$) can be ignored,
this equation alone governs the bulk-flow-mediated mode dynamics.
Equation (\ref{eq:NormElec}), a combination of the parallel-induction
and parallel-vorticity equations, governs the electron dynamics. The
left side of this equation is the contribution from the diffusion
of the parallel field and third term on the right side is the contribution
from the Hall term ($\hat{k}_{\parallel}\hat{d}_{i}\hat{B_{x}^{\prime\prime}}$).
The $\bar{\tau}$ parameters scale as $\beta^{-1}$ and are typically
important only at small values of $\beta$. The dominant $\beta^{-1}$
contributions result from the gradient of the electron pressure in
Ohm's law (terms involving $\hat{\omega}_{*n}$) and perpendicular
compressibility (otherwise). There are other contributions to $\bar{\tau}_{Q}$
and $\bar{\tau}_{\xi}$ from the parallel-field inertia and the $\mathbf{v}\times\mathbf{B}$
term, respectively, however these term are unimportant from a practical
perspective. The first and last term on the right side are also contributions
from perpendicular compressibility and are important in the moderate-$\beta$
transition regime (PR2).

\section{Drift-Tearing Dispersion Relations by Parametric Regime \label{sec:Dispersion}}

\noindent Once solutions for $\bar{Q}$ and $\bar{\xi}$ are found,
the dispersion relation may be computed by integrating the radial
induction equation (Eqn.~(\ref{eq:linRadInd})) and applying the
boundary condition $\tilde{B}_{r}^{\prime}\left(\pm\infty\right)=0$.
The resulting equation is
\begin{equation}
D=\int_{-\infty}^{\infty}d\bar{x}\left(1-\bar{x}\bar{\xi}-\bar{x}\bar{Q}\right)=\frac{\hat{k_{\parallel}^{\prime}}^{1/2}\hat{\Delta^{\prime}}}{\hat{\gamma}_{e}\hat{\gamma}_{ExB}^{1/4}S_{g}^{3/4}}\label{eq:DispRelInt}
\end{equation}
where we have defined $D$ for notational convenience. The right side
of this expression is the contribution from resistivity, thus the
integrand of left side of this expression is the ideal radial Ohm's
law. As resistivity is only significant in the layer, proper matching
of the inner and outer region solutions ensures the integrand vanishes
outside the layer and the integral converges.

\begin{table}
\begin{centering}
\begin{tabular}{ccccccc}
\hline 
 & PR1a & PR2 & PR3 & PR4 & PR5 & PR1b\tabularnewline
\hline 
\hline 
Regime & $\bar{\sigma}^{2}<<1$  & $\bar{\sigma}^{2}\sim1$  & $\bar{\sigma}^{2}>>1$  & $\bar{\sigma}^{2}>>1$  & $\bar{\sigma}^{2}>>1$  & $\bar{\tau}_{Q}>>\bar{\sigma}^{2}$ \tabularnewline
boundary & and & and & and & and & and & and\tabularnewline
 & $\bar{\tau}_{Q}<<1$ & $\bar{\tau}_{Q}<<1$ & $\bar{\tau}_{Q}<<\bar{\sigma}$ & $\bar{\tau}_{Q}\sim\bar{\sigma}$ & $\bar{\sigma}<<\bar{\tau}_{Q}<<\bar{\sigma}^{2}$ & $\bar{\tau}_{Q}>>1$ \tabularnewline
 &  & or &  &  &  & \tabularnewline
 &  & $\bar{\tau}_{Q}<<\bar{\sigma}$ &  &  &  & \tabularnewline
Dominant field & $\bar{\xi}$ & $\bar{\xi}$ and $\bar{Q}$ & $\bar{Q}$ & $\bar{Q}$ & $\bar{Q}$ & $\bar{\xi}$ and $\bar{Q}$\tabularnewline
\hline 
$B_{x}$ diffusion & \CheckmarkBold{} & \CheckmarkBold{} & \CheckmarkBold{} & \CheckmarkBold{} & \CheckmarkBold{} & \CheckmarkBold{}\tabularnewline
$B_{\parallel}$ diffusion &  & \CheckmarkBold{} & \CheckmarkBold{} & \CheckmarkBold{} &  & \tabularnewline
Hall decoupling &  & \CheckmarkBold{} & \CheckmarkBold{} & \CheckmarkBold{} & \CheckmarkBold{} & \tabularnewline
$\nabla\cdot\mathbf{v}$ decoupling &  & \CheckmarkBold{} &  & \CheckmarkBold{} & \CheckmarkBold{} & \tabularnewline
\hline 
no drift reference & \cite{Furth_63_PoF_459} & \cite{Ahedo_09_PPaCF_55018}  & \cite{Bulanov_92_PoFB_2499}  & \cite{Mirnov_04_PoP_4468} & \cite{Drake_77_PoF_1341} & \cite{Furth_63_PoF_459}\tabularnewline
drift reference & new & new & new & new & \cite{Drake_77_PoF_1341} & \cite{Coppi_64_PoF_1501}\tabularnewline
\hline 
Hall drift  & \CheckmarkBold{} & \CheckmarkBold{} & \CheckmarkBold{} & \CheckmarkBold{} & \CheckmarkBold{} & \CheckmarkBold{}\tabularnewline
$\nabla p_{e}$ drift  &  &  &  & \CheckmarkBold{} & \CheckmarkBold{} & \CheckmarkBold{}\tabularnewline
$\nabla\cdot\mathbf{v}$ drift &  & \CheckmarkBold{} &  & \CheckmarkBold{} & \CheckmarkBold{} & \CheckmarkBold{}\tabularnewline
\hline 
\end{tabular}
\par\end{centering}

\caption{A summary of the parametric regime boundaries, significant terms and
fields, and prior references if applicable.\label{tab:regimes}}

\end{table}

We next derive the dispersion relation in the various parametric regimes
as summarized in Tab.~\ref{tab:regimes}. We begin in the single-fluid
regime (PR1) with $\bar{\tau}_{Q}<<1$ (near PR2) and work our way
clockwise around Fig.~\ref{fig:paramSpace}. We do not address PR6,
which was solved numerically in Ref.~\cite{Ahedo_09_PPaCF_55018},
as it is of limited relevance to fusion-plasma experiments. We finish
again in the single-fluid regime (PR1) with $\bar{\tau}_{Q}>>\bar{\sigma}^{2}$
(near PR6) where we recover the drift-tearing result of Ref.~\cite{Coppi_64_PoF_1501}.

\subsection{PR1a}

We use PR1a as a notation for the upper left quadrant of Fig.~\ref{fig:paramSpace}
where 
\begin{equation}
\bar{\tau}_{Q},\;\bar{\tau}_{\xi},\;\bar{\tau}_{B},\;\bar{\sigma}^{2}\;,\bar{\Lambda}<<1\sim\bar{x}\sim\bar{\xi}\;.
\end{equation}
Examination of the system of tearing equations (Eqns.~(\ref{eq:NormIon})
and (\ref{eq:NormElec})) shows $\bar{Q}<<\bar{\xi}$. Thus the electron
equation (Eqn.~(\ref{eq:NormElec})) may be ignored and the governing
equation is simply
\begin{equation}
\bar{R}\bar{\xi}^{\prime\prime}=\bar{x}^{2}\bar{\xi}-\bar{x}\;.
\end{equation}
The solution for $\bar{\xi}$ can be expressed in terms of the parabolic
cylinder function,
\begin{equation}
U\left(0,\bar{x}\right)=\frac{\bar{x}}{2}\int_{0}^{1}d\mu\left(1-\mu^{2}\right)^{-1/4}\exp\left[-\frac{\mu\bar{x}^{2}}{2}\right]\;,
\end{equation}
as $\bar{\xi}=\bar{R}^{-1/4}U\left(0,\bar{R}^{-1/4}\bar{x}\right)$.
Integrating Eqn.~(\ref{eq:DispRelInt}), the drift dispersion relation
is 
\begin{equation}
\hat{\gamma}_{e}\hat{\gamma}_{gvi}^{1/4}=\hat{\gamma}_{MHD}^{5/4}\label{eq:DispPR1a}
\end{equation}
 where $\hat{\gamma}_{MHD}$ is the single-fluid growth rate without
drift effects, 
\begin{equation}
\hat{\gamma}_{MHD}=S_{g}^{-3/5}\left(\frac{\hat{\Delta^{\prime}}}{\sqrt{2}\Gamma\left(\nicefrac{3}{4}\right)^{2}}\right)^{4/5}\hat{k_{\parallel}^{\prime}}^{2/5}\;.
\end{equation}

\subsection{PR2}

Regime PR2 is the transition at moderate $\beta$ between the single-fluid
regime, PR1, and the electron-MHD regime, PR3. Here we assume $\bar{\Lambda}\sim\bar{x}\sim1$,
$\bar{\xi}\sim\bar{Q}$ and
\begin{equation}
\bar{\tau}_{Q},\;\bar{\tau}_{\xi},\;\bar{\tau}_{B}<<1\quad or\quad\bar{\tau}_{Q},\;\bar{\tau}_{\xi},\;\bar{\tau}_{B}<<\bar{\sigma}\;.
\end{equation}
Thus the system of tearing equations becomes
\begin{equation}
\bar{Q}^{\prime\prime}=\bar{R}^{-1}\bar{x}^{2}\bar{Q}+\bar{R}\bar{\sigma}^{2}\bar{\xi}^{\prime\prime}+\bar{\Lambda}\bar{x}\;,
\end{equation}
and
\begin{equation}
\bar{R}\bar{\xi}^{\prime\prime}=\bar{x}^{2}\left(\bar{Q}+\bar{\xi}\right)-\bar{x}\;.
\end{equation}
Following the method outlined in Ref.~\cite{Ahedo_09_PPaCF_55018}
for the solution of a similar system of equations (where $\bar{R}\rightarrow1$
and $\bar{\Lambda}\rightarrow0$), we transform this system of equations
into two independent parabolic cylinder equations, 
\begin{equation}
\lambda_{i}^{-1}\bar{V}_{i}^{\prime\prime}=\bar{x}^{2}\bar{V}_{i}-C_{i}\bar{x}\;,
\end{equation}
where $\bar{V}_{i}=\bar{\xi}+a_{i}\bar{Q}$ and $i=1,2$. This transformation
requires 
\begin{equation}
\lambda_{i}=\bar{R}^{-1}+\bar{\sigma}^{2}a_{i}\;,
\end{equation}
\begin{equation}
a_{i}=\frac{1}{2}\pm\frac{1}{2}\sqrt{1+\frac{4}{\bar{R}\bar{\sigma}^{2}}}\;,
\end{equation}
and
\begin{equation}
C_{i}=1-\bar{\Lambda}\bar{R}\left(a_{i}-1\right)\;.
\end{equation}
The solution for each $\bar{V}_{i}$ is $\bar{V}_{i}=\lambda_{i}^{1/4}C_{i}U\left(0,\lambda_{i}^{1/4}\bar{x}\right)$.
Integrating Eqn.~(\ref{eq:DispRelInt}) to find the dispersion relation
gives
\begin{equation}
D=\frac{2\pi\Gamma\left(\nicefrac{3}{4}\right)}{\Gamma\left(\nicefrac{1}{4}\right)}\left[\frac{C_{1}a_{1}\lambda_{1}^{-\nicefrac{1}{4}}-C_{2}a_{2}\lambda_{2}^{-\nicefrac{1}{4}}}{a_{1}-a_{2}}\right]\;.\label{eq:DispPR2}
\end{equation}
This may be expressed in a more explicit form as $D=\sqrt{2}\Gamma\left(\nicefrac{3}{4}\right)^{2}f_{2}\left(\bar{\sigma},\bar{R},\bar{\Lambda}\right)$,
where
\begin{multline}
f_{2}\left(\bar{\sigma},\bar{R},\bar{\Lambda}\right)=\underset{i=1,2}{\sum}\frac{1}{2}\left[1-\frac{\bar{\Lambda}\bar{R}}{2}\left(\left(-1\right)^{i}\sqrt{1+\frac{4}{\bar{R}\bar{\sigma}^{2}}}-1\right)\right]\left[1+\left(-1\right)^{i}\left(1+\frac{4}{\bar{R}\bar{\sigma}^{2}}\right)^{-\nicefrac{1}{2}}\right]\\
\times\left[\bar{R}^{-1}+\frac{\bar{\sigma}^{2}}{2}+\left(-1\right)^{i}\bar{\sigma}\sqrt{\frac{\bar{\sigma}^{2}}{4}+\bar{R}^{-1}}\right]^{-\nicefrac{1}{4}}\;.
\end{multline}
The limits of this expression under the same approximations as PR1a
and PR3 are consistent with the dispersion relations found in these
regimes. Consider the limit where $\bar{\sigma}^{2}<<1$, in this
case $f_{2}\left(\bar{\sigma},\bar{R},\bar{\Lambda}\right)\rightarrow\left(1+\bar{\Lambda}\bar{R}/4\right)\bar{R}^{1/4}$.
With the additional limit $\bar{\Lambda}<<1$ (as is the case in PR1a),
$f_{2}\left(\bar{\sigma},\bar{R},\bar{\Lambda}\right)\rightarrow\bar{R}^{1/4}$
and we recover Eqn.~(\ref{eq:DispPR1a}). In the limit where $\bar{\sigma}^{2}>>1$,
$f_{2}\left(\bar{\sigma},\bar{R},\bar{\Lambda}\right)\rightarrow\bar{\sigma}^{-1/2}$.
As we shall see in the next subsection, this limit is the dispersion
relation found in the electron-MHD regime, PR3.

\subsection{PR3}

In the electron-MHD regime, the resistive diffusion of $B_{\parallel}$
balances the Hall term in the parallel induction equation, and the
parallel-vorticity equation is not needed. The orderings of this regime
are a small tearing layer and large $B_{\parallel}$, $\bar{x}^{-1}\sim\bar{\sigma}^{1/2}\sim\bar{Q}$
, small ion displacement, $\bar{\xi}\sim\bar{\sigma}^{-3/2}$, and
large $d_{i}$, $\bar{\sigma}^{2}>>1$, such that $\bar{\Lambda}<<\bar{\sigma}^{2}$,
$\bar{\tau}_{\xi}<<\bar{\sigma}^{3}$, $\bar{\tau}_{B}<<\bar{\sigma}^{3/2}$
and $\bar{\tau}_{Q}<<\bar{\sigma}$. After substituting the ordered
electron equation, Eqn.~(\ref{eq:NormElec}), into the ordered ion
equation, Eqn.~(\ref{eq:NormIon}), the governing equation in this
regime is
\begin{equation}
\bar{Q}^{\prime\prime}=\bar{\sigma}^{2}\bar{x}^{2}\bar{Q}-\bar{\sigma}^{2}\bar{x}\;.
\end{equation}
The solution to this equation is $\bar{Q}=\sqrt{\bar{\sigma}}U\left(0,\sqrt{\bar{\sigma}}\bar{x}\right)$.
Integrating Eqn.~(\ref{eq:DispRelInt}), we find $D=\sqrt{2}\Gamma\left(\nicefrac{3}{4}\right)^{2}\bar{\sigma}^{-1/2}$
(the limit of $D$ from PR2 when $\bar{\sigma}^{2}>>1$) and the dispersion
relation is then
\begin{equation}
\hat{\gamma}_{e}=S_{g}^{-1/2}\left(\hat{d}_{i}\hat{k_{\parallel}^{\prime}}\right)^{1/2}\frac{\hat{\Delta^{\prime}}}{\sqrt{2}\Gamma\left(\nicefrac{3}{4}\right)^{2}}\;.\label{eq:DispPR3}
\end{equation}
In this regime the growth rate scales as $d_{i}^{1/2}S^{-1/2}$ and
the mode simply rotates at the electron drift frequency; there is
no drift stabilization. This result is not particularly surprising,
as the mode is mediated purely by the electron fluid through the induction
equation. Contributions from ion compressibility, parallel ion flows
and ion vorticity do not play a role.

\subsection{PR4\label{sub:PR4}}

The PR4 regime is the transition between the $B_{\parallel}$-diffusion
(PR3) and the semicollisional (PR5) regimes. The orderings of this
regime are similar to PR3; a small tearing layer with a large $B_{\parallel}$,
$\bar{x}^{-1}\sim\bar{\sigma}^{1/2}\sim\bar{Q}$, small ion displacement,
$\bar{\xi}\sim\bar{\sigma}^{-3/2}$, however $\bar{\tau}_{Q}$ is
comparable to the normalized ion skin depth which is large, $\bar{\sigma}^{2}>>1$,
such that $\bar{\Lambda}<<\bar{\sigma}^{2}$, $\bar{\tau}_{\xi}<<\bar{\sigma}^{3}$,
$\bar{\tau}_{B}\sim\bar{\sigma}^{3/2}$ and $\bar{\tau}_{Q}\sim\bar{\sigma}$.
Thus the $\bar{\tau}_{Q}$ and $\bar{\tau}_{B}$ contributions must
both be retained in Eqn.~(\ref{eq:NormElec}), and the system of
tearing equations becomes
\begin{equation}
\bar{Q}^{\prime\prime}=\bar{\tau}_{Q}\bar{Q}+\bar{R}\bar{\sigma}^{2}\bar{\xi}^{\prime\prime}-\bar{\tau}_{B}\;,
\end{equation}
and
\begin{equation}
\bar{R}\bar{\sigma}^{2}\bar{\xi}^{\prime\prime}=\bar{\sigma}^{2}\bar{x}^{2}\bar{Q}-\bar{\sigma}^{2}\bar{x}\;.
\end{equation}
These equations may be combined into a single non-homogeneous parabolic
cylinder equation for $\bar{Q}$,
\begin{equation}
\bar{Q}^{\prime\prime}=\bar{\tau}_{Q}\bar{Q}+\bar{\sigma}^{2}\bar{x}^{2}\bar{Q}-\bar{\sigma}^{2}\bar{x}-\bar{\tau}_{B}\;.
\end{equation}
The solution to this equation up to a constant of integration, following
the method outlined in Ref.~\cite{hazeltineMeiss}, is
\begin{equation}
\bar{Q}\left(\bar{x}\right)=A_{+}\int_{0}^{\infty}\exp\left(ik\bar{x}\sqrt{\frac{\bar{\sigma}}{2}}\right)\frac{U\left(a,k\right)}{U\left(a,0\right)}dk+A_{-}\int_{\infty}^{0^{-}}\exp\left(ik\bar{x}\sqrt{\frac{\bar{\sigma}}{2}}\right)\frac{U\left(a,k\right)}{U\left(a,0\right)}dk\;,
\end{equation}
with the constraint
\begin{equation}
A_{+}-A_{-}=\frac{i\bar{\sigma}^{3/2}}{\sqrt{2}}+\frac{\bar{\tau}_{B}}{2\bar{\sigma}}\frac{U\left(a,0\right)}{U^{\prime}\left(a,0\right)}\;,
\end{equation}
where $a=\bar{\tau}_{Q}/2\bar{\sigma}$. The constant of integration
(either $A_{+}$ or $A_{-}$) is found by matching the layer equations
with the outer solution (in practice, requiring that the integral
of Eqn.~(\ref{eq:DispRelInt}) converges), which provides the additional
condition $A_{+}=-i\sqrt{\bar{\sigma}/2}$. Integrating Eqn.~(\ref{eq:DispRelInt})
determines the dispersion relation as 
\begin{equation}
\hat{\gamma}_{e}\frac{\Gamma\left[\left(3+\bar{\tau}_{Q}/\bar{\sigma}\right)/4\right]}{\Gamma\left[\left(1+\bar{\tau}_{Q}/\bar{\sigma}\right)/4\right]}=S_{g}^{-1/2}\sqrt{\hat{k_{\parallel}^{\prime}}\hat{d}_{i}}\frac{\hat{\Delta^{\prime}}}{2\pi}\;.\label{eq:DispPR4}
\end{equation}
Drift effects modify the dispersion relation through the left side,
in particular the drift modified growth rate $\hat{\gamma}_{e}$ and
the drift effects contained in $\bar{\tau}_{Q}$ and $\bar{\sigma}$.
In the limit where $\bar{\tau}_{Q}<<\bar{\sigma}$, the left side
of Eqn.~(\ref{eq:DispPR4}) becomes $\hat{\gamma}_{e}\Gamma\left(\nicefrac{3}{4}\right)^{2}/\sqrt{2}\pi$,
consistent with the dispersion relation of PR3. In the opposite limit,
where $\bar{\tau}_{Q}>>\bar{\sigma}$ the left side of the equation
becomes $\hat{\gamma}_{e}\sqrt{\bar{\tau}_{Q}}/2\sqrt{\bar{\sigma}}$,
which is consistent with the dispersion relation found in the next
section for the semicollisional regime, PR5. Although $\bar{\tau}_{B}$,
which scales similarly in magnitude to $\bar{\tau}_{Q}$, affects
the eigenfunction, it does not modify the growth rate. In the limit
of PR3, both $\bar{\tau}_{B}$ and $\bar{\tau}_{Q}$ are small and
thus the results are consistent. In the limit of PR5, where $\bar{\tau}_{B}$
is again expected to be large, $\bar{\tau}_{B}$ contributes an even
parity term to the eigenfunction and thus again does not contribute
to the dispersion relation after integration of Eqn.~(\ref{eq:DispRelInt}).

\subsection{PR5\label{sub:PR5}}

The orderings in the semicollisional regime are similar to PR3 and
PR4, with a large $B_{\parallel}$, $\bar{x}^{-1}\sim\bar{\sigma}^{1/2}\sim\bar{Q}$,
and small ion displacement, $\bar{\xi}\sim\bar{\sigma}^{-3/2}$. However
in this regime the $\bar{\tau}$ terms are larger than normalized
ion skin depth (but not too large), $\bar{\sigma}^{2}>>1$, such that
$\bar{\Lambda}<<\bar{\sigma}^{2}$, $\bar{\tau}_{\xi}<<\bar{\sigma}^{3}$,
and $\bar{\sigma}<<\bar{\tau}_{Q}\sim\bar{\tau}_{B}<<\bar{\sigma}^{2}$.
The diffusion of $B_{\parallel}$ may be neglected and the Hall term
in Eqn.~(\ref{eq:NormElec}) is balanced by the $\bar{\tau}_{Q}$
and $\bar{\tau}_{B}$ terms. After substitution of the ordered electron
equation, Eqn.~(\ref{eq:NormElec}), into the ordered ion equation,
Eqn.~(\ref{eq:NormIon}), the governing equation for this regime
is
\begin{equation}
0=\bar{\tau}_{Q}\bar{Q}+\bar{\sigma}^{2}\bar{x}^{2}\bar{Q}-\bar{\sigma}^{2}\bar{x}-\bar{\tau}_{B}\;.
\end{equation}
The solution is algebraic,
\begin{equation}
\bar{Q}=\frac{\bar{\sigma}^{2}\bar{x}+\bar{\tau}_{B}}{\bar{\sigma}^{2}\bar{x}^{2}+\bar{\tau}_{Q}}\;.
\end{equation}
The dispersion relation, found by integrating Eqn.~(\ref{eq:DispRelInt}),
is then
\begin{equation}
\hat{\gamma}_{e}\hat{\tau}_{Q}^{1/2}=S_{g}^{-1/2}\frac{\hat{k_{\parallel}^{\prime}}\hat{\Delta^{\prime}}}{\pi}\hat{d}_{i}\;.\label{eq:DispPR5}
\end{equation}
The growth rate scales as $\rho_{s}^{2/3}S_{g}^{-1/3}$ and drift
effects are contained on the left side of Eqn.~(\ref{eq:DispPR5}).
The eigenfunction, $\bar{Q}$, only contains even-parity contributions
from $\bar{\tau}_{B}$, which vanish during the integration of Eqn.~(\ref{eq:DispRelInt})
and thus do not contribute to the dispersion relation. 

This is the two-fluid drift-regime first described by Drake and Lee
\cite{Drake_77_PoF_1341}. Simplifying this expression further by
assuming $\beta<<1$ (which defines this regime), $\sigma_{pe}=\sigma_{qi}=-\sigma_{qe}=1$,
we find
\begin{equation}
\hat{\gamma}_{e}^{2/3}\left(\frac{\hat{\gamma}_{ExB}\hat{\gamma}_{i}}{f_{Ti}\hat{\gamma}_{e}+f_{Te}\hat{\gamma}_{i}}\right)^{1/3}=S_{g}^{-1/3}\left(\frac{\hat{k_{\parallel}^{\prime}}\hat{\Delta^{\prime}}}{\pi}\hat{c}_{s}\hat{d}_{i}\right)^{2/3}\;.
\end{equation}
When the electron temperature is much larger than the ion, $f_{Ti}=0$
and $f_{Te}=1$, the standard two-thirds, one-third dispersion relation
is attained. Our results are not identical to Drake and Lee; however
we do not include ion and electron gyroviscosity or heat-flux contributions
to the frictional force. The inclusion of cross heat flux cancels
contributions from the pure density-gradient drifts in the dispersion
relation, as with $\sigma_{q\alpha}=0$ but $\sigma_{pe}=1$, one
instead finds
\begin{equation}
\hat{\gamma}_{e}^{2/3}\left(\frac{\left(\hat{\gamma}_{ExB}-f_{Te}\Gamma i\hat{\omega}_{*n}\right)\left(\hat{\gamma}_{ExB}+f_{Ti}\Gamma i\hat{\omega}_{*n}\right)}{\hat{\gamma}_{e}}\right)^{1/3}=S_{g}^{-1/3}\left(\frac{\hat{k_{\parallel}^{\prime}}\hat{\Delta^{\prime}}}{\pi}\hat{c}_{s}\hat{d}_{i}\right)^{2/3}\;.
\end{equation}

\subsection{PR1b}

This regime is the low-$\beta$, drift limit of the single-fluid regime.
With the corresponding orderings, $1<<\bar{\tau}_{Q}\sim\bar{\tau}_{\xi}$,
$\bar{\sigma}^{2}<<\bar{\tau}_{Q}\sim\bar{\tau}_{\xi}$ and $\bar{\tau}_{B}\sim\bar{\Lambda}\sim1$,
the $\bar{\tau}_{Q}$ and $\bar{\tau}_{\xi}$ terms balance in the
electron equation, Eqn.~(\ref{eq:NormElec}), and thus $\bar{\xi}\sim\bar{Q}$
as $\bar{\tau}_{Q}\bar{Q}=-\bar{\tau}_{\xi}\bar{\xi}$. Substituting
this balance into Eqn.~(\ref{eq:NormIon}) the governing equation
becomes
\begin{equation}
\bar{R}\bar{\xi}^{\prime\prime}=\bar{x}^{2}\left(1-\frac{\bar{\tau}_{\xi}}{\bar{\tau}_{Q}}\right)\bar{\xi}-\bar{x}\;.
\end{equation}
The solution is $\bar{\xi}=\bar{R}^{-1/4}M^{-3/4}U\left(0,\bar{R}^{-1/4}M^{1/4}\bar{x}\right)$
where $M=1-\bar{\tau}_{\xi}/\bar{\tau}_{Q}$. Assuming $\beta<<1$,
and thus $M\simeq\hat{\gamma}_{e}/\hat{\gamma}_{i}$, integration
of Eqn.~(\ref{eq:DispRelInt}) gives the dispersion relation, 
\begin{equation}
\hat{\gamma}_{MHD}^{5/4}=\hat{\gamma}_{e}^{3/4}\hat{\gamma}_{i}^{1/4}\hat{\gamma}_{gvi}^{1/4}\;.
\end{equation}
With an exact gyroviscous cancellation, $\hat{\gamma}_{gvi}=\hat{\gamma}_{ExB}$,
this is the standard drift-tearing dispersion relation as found by
Coppi \cite{Coppi_64_PoF_1501}.

\section{Verification of the NIMROD Code \label{sec:Verification}}

\noindent We may now use the dispersion relations of Sec.~\ref{sec:Dispersion}
to verify the implementation of the unreduced extended-MHD equations
in the initial-value NIMROD code \cite{Sovinec_04_JoCP_355}. NIMROD
is primarily designed as a nonlinear-physics code. However, it uses
the linear response of the perturbed system as a preconditioner during
nonlinear solves. This functionality makes available the linearized
equation within NIMROD and thus permits our verification exercise.
This verification is a partial test of the NIMROD equation implementation
as well as a test of the time and spatial discretizations.

Cases are implemented as a periodic-in-y, symmetric-in-z box within
NIMROD. Each specific equilibrium is generated by specifying the equilibrium
magnetic-shear-scale length ($L_{S}$), the ratio of the magnetic
shear to guide field ($\epsilon_{B}$), the plasma $\beta$, the equilibrium
pressure-gradient-scale length ($L_{P}$), and the ratio of the sheared
to background pressure ($\epsilon_{P}$). Equilibrium fields are computed
by solving the MHD-force balance equations based on a hyperbolic-secant-squared
parallel-current profile, 
\begin{equation}
\lambda_{0}=\mu_{0}\frac{\mathbf{J}_{0}\cdot\mathbf{B}_{0}}{B_{0}^{2}}=\frac{\epsilon_{B}}{L_{S}}sech^{2}\left(\frac{x}{L_{S}}\right)\;,
\end{equation}
and a hyperbolic-tangent pressure profile,
\begin{equation}
\Gamma\mu_{0}\frac{p_{0}}{B_{0}^{2}}=\beta\left(1+\epsilon_{P}\tanh\left(\frac{x}{L_{P}}\right)\right)\;.
\end{equation}
Our cases use comparable magnetic-shear and pressure-gradient scale
lengths, $L_{s}=L_{p}$, and impose this gradient with a dominant
density profile to avoid ITG-like modes (see Ref.~\cite{Schnack_13_PoP}).
The fraction of the pressure gradient that results from the density
profile, $f_{n}=n_{0}^{\prime}p_{0}/n_{0}p_{0}^{\prime}$, always
equals or exceeds $1/2$. Drift effects are included when $\epsilon_{P}\neq0$.
For cases with $\epsilon_{P}=0$, the tearing stability parameter,
$\Delta^{\prime}$, may be computed analytically for this equilibrium
(Ref.~\cite{Ahedo_09_PPaCF_55018}):
\begin{equation}
\Delta^{\prime}=\frac{2}{L_{S}}\left(\frac{1}{kL_{S}}-kL_{S}\right)\;.\label{eq:SlabDeltaPrime}
\end{equation}
For cases with $\epsilon_{P}\neq0$, we use NIMROD to infer that $\Delta^{\prime}$
is unchanged. As $p_{0}^{\prime}$ is increased, if the growth rate
from NIMROD computations with a single-fluid model is unchanged then
$\Delta^{\prime}$ is constant. Equation (\ref{eq:SlabDeltaPrime})
assumes an infinite-in-x domain. This is, of course, not practical
for the NIMROD finite-element computations where instead a large ratio
of $D_{x}/L_{s}$ is used to approximate the infinite domain, where
$D_{x}$ is the box half length in the x dimension. Our cases use
$D_{x}/L_{s}=6$ with a 96 radial bi-cubic elements packed near the
resonant surface where the single-fluid growth rate discrepancy between
NIMROD and the analytics is less than $1\%$.

Table \ref{tab:VerficationCases} summarizes the parameters used for
our verification studies. In a practical verification exercise, the
physical parameter space (equilibrium characteristic values, length
scales and gradients) affect the derived parameter space (Lundquist
number, tearing stability parameter, ion skin depth, $\beta$ and
drift frequencies) in a complex manner. The locations of these cases
in the $\bar{\sigma}-\bar{\tau}_{Q}$ parameter space in the limit
where $\hat{\omega}_{*}\rightarrow0$ is superimposed onto Fig.~\ref{fig:paramSpace}.
In general as $\hat{\omega}_{*}$ increases, $\bar{\sigma}$  marginally
increases and the $\bar{\tau}$ parameters increase linearly moving
the cases down (and slightly to the right) in the $\bar{\sigma}-\bar{\tau}_{Q}$
parameter space of Fig.~\ref{fig:paramSpace}, as illustrated with
dashed lines. Our choice of scan locations in the $\bar{\tau}_{Q}$-$\bar{\sigma}$
phase space is the result of a combination of finding a representative
sample of cases to fill the experimentally relevant parameter space
of Table \ref{tab:experimentalParam}, choosing cases which are able
to achieve reasonable $\hat{\omega}_{*}\propto\epsilon_{P}\beta\hat{d}_{i}$
with $\epsilon_{P}<1$ (which avoids negative pressure regions), and
testing the analytics in a variety of regimes. 

\begin{table}
\begin{centering}
\begin{tabular}{cccccccccc}
\hline 
case & $k_{\perp}d_{i}$ & $\beta$ & $\bar{\sigma}$ & $\bar{\tau}_{Q}$ & $\bar{\tau}_{B}$ & $\bar{\tau}_{\xi}$ & $\bar{\Lambda}$ & regime & stabilization\tabularnewline
\hline 
\hline 
A (Fig.~\ref{fig:scan2-3}) & $0.002$ & $0.1$ & 0.027 & 0.056 & $1.72\times10^{-4}$ & 0.079 & 0.99 & PR2/PR0 & weak/strong \tabularnewline
B (Fig.~\ref{fig:scan6-2}) & $0.064$ & $0.1$ & 0.91 & 0.065 & 0.0025 & 0.088 & 1.2 & PR2/PR0 & strong \tabularnewline
C (Fig.~\ref{fig:scan3-5}) & $2.048$ & $0.1$ & 65 & 0.37 & 0.061 & 0.29 & 066 & PR3/PR4 & none\tabularnewline
D (Fig.~\ref{fig:scan4-2}) & $2.048$ & $1.56\times10^{-3}$ & 54 & 22 & 0.017 & 18 & 0.87 & PR4-PR6 & moderate\tabularnewline
\hline 
\end{tabular}
\par\end{centering}

\caption{Parameters for the verification $\omega_{*}$ scans with the NIMROD
code. The normalized parameters are evaluated at $\hat{\omega}_{*}=1.05\times10^{-5}$
and are modified by drift effects. All scans use $\sigma_{pe}=\sigma_{qi}=-\sigma_{qe}=1$,
$S=3.5\times10^{7}$, $\hat{\Delta}^{\prime}=1.46$, $kL_{s}=0.76$
and $\epsilon_{B}=0.02$. Cases use the electron-to-ion-mass ratio
from a Deuterium gas discharge of $2.7\times10^{-4}$ unless otherwise
mentioned. \label{tab:VerficationCases}}
\end{table}

All cases rotate in the electron diamagnetic direction. The dominant
$\omega_{*e}$ influence results from the denominator of the right
side of Eqn.~(\ref{eq:DispRelInt}). In the electron-MHD regime of
PR3, where the ion dynamics no longer influence the mode, the mode
is at rest in the frame of the electron fluid. 

\begin{figure}
\begin{centering}
\includegraphics[width=0.45\paperwidth]{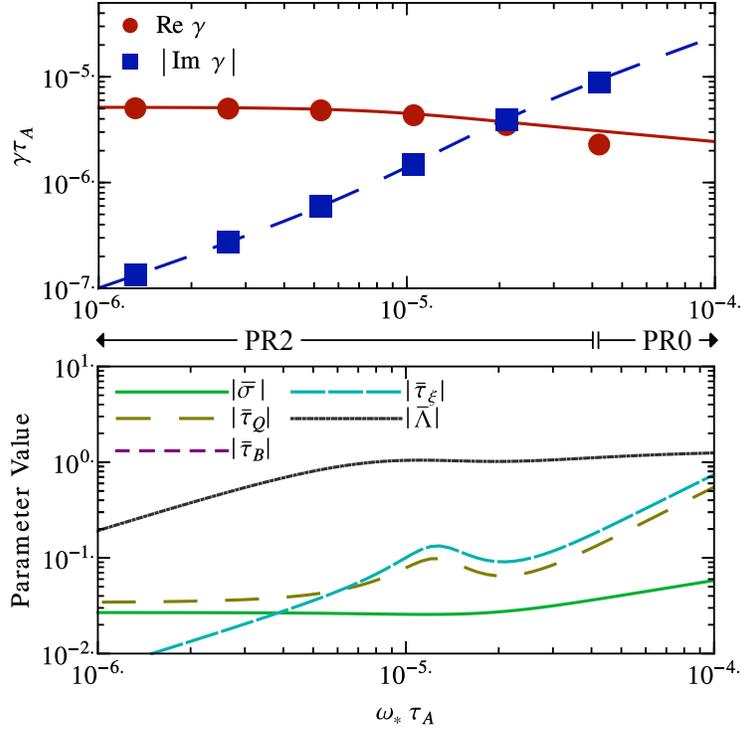}
\par\end{centering}

\caption{Scan A growth rates and normalized parameters with $\mathbf{v}_{E\times B}=-\mathbf{v}_{\nabla p}$
and $f_{n}=1$. The converged results from NIMROD runs (points) are
compared with the drift analytics of PR2 (lines, Eqns.~(\ref{eq:DispRelInt})
and (\ref{eq:DispPR2})). \label{fig:scan2-3}}
\end{figure}

The scan A growth-rate comparison at moderate $\beta$ (0.1) and low
$\hat{d}_{i}$ (0.002) between NIMROD runs and the dispersion relation
of PR2 is shown in Fig.~\ref{fig:scan2-3}. Good agreement is achieved
until $\bar{\tau}_{Q}\sim1$ (the five left-most points in the figure
agree with the analytics with less than a $3\%$ error) and the mode
enters the regime of PR0. Although there are no analytics for this
regime, we note NIMROD predicts stronger drift-stabilization in PR0
than the relatively weak effect predicted by the drift analytics in
PR2. In fact, at larger values of $\hat{\omega}_{*}$ NIMROD predicts
complete stabilization in PR0 as NIMROD cases at $\hat{\omega}_{*}=7.7\times10^{-5}$
are stable.

\begin{figure}
\begin{centering}
\includegraphics[width=0.45\paperwidth]{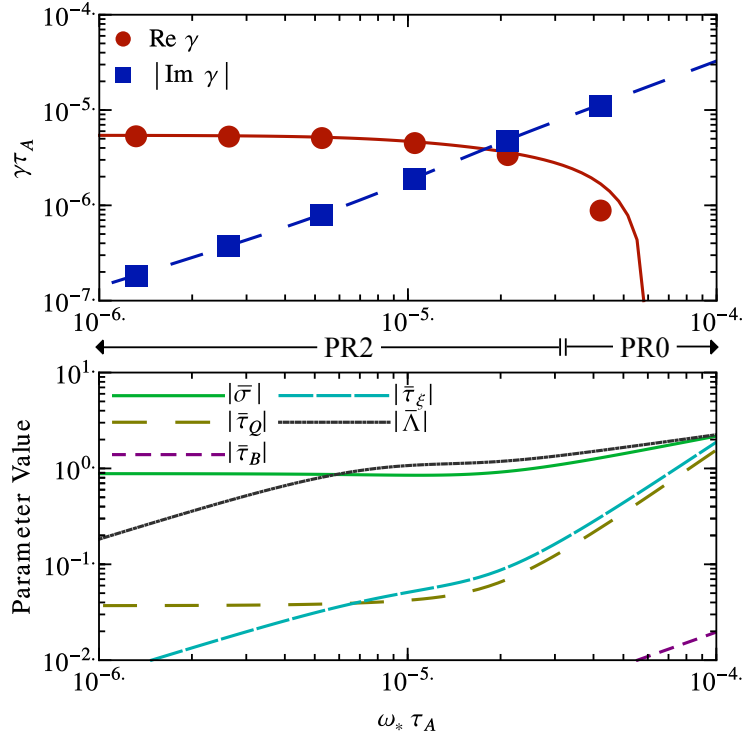}
\par\end{centering}

\caption{Scan B growth rates and normalized parameters with $\mathbf{v}_{E\times B}=0$
and $f_{n}=0.5$. The converged results from NIMROD runs (points)
are compared with the drift analytics of PR2 (lines, Eqns.~(\ref{eq:DispRelInt})
and (\ref{eq:DispPR2})). \label{fig:scan6-2}}
\end{figure}

Figure \ref{fig:scan6-2} shows the scan B result of a verification
scan at moderate $\hat{d}_{i}$ (0.064) and $\beta$ (0.1) which again
begins in PR2 and transitions to PR0. Similar to scan A, as $\hat{\omega}_{*}$
is increased $\bar{\tau}_{Q}$ approaches unity and the mode ultimately
enters PR0 where there is no analytic solution. However, unlike scan
A, both the computations and the analytics predict complete stabilization
of the mode at qualitatively similar values of $\omega_{*}$ (NIMROD
computations at $\hat{\omega}_{*}=5.2\times10^{-5}$ are stable).
Similar to scan A, the first five left-most points are within 3\%
of the analytic results.

\begin{figure}
\begin{centering}
\includegraphics[width=0.45\paperwidth]{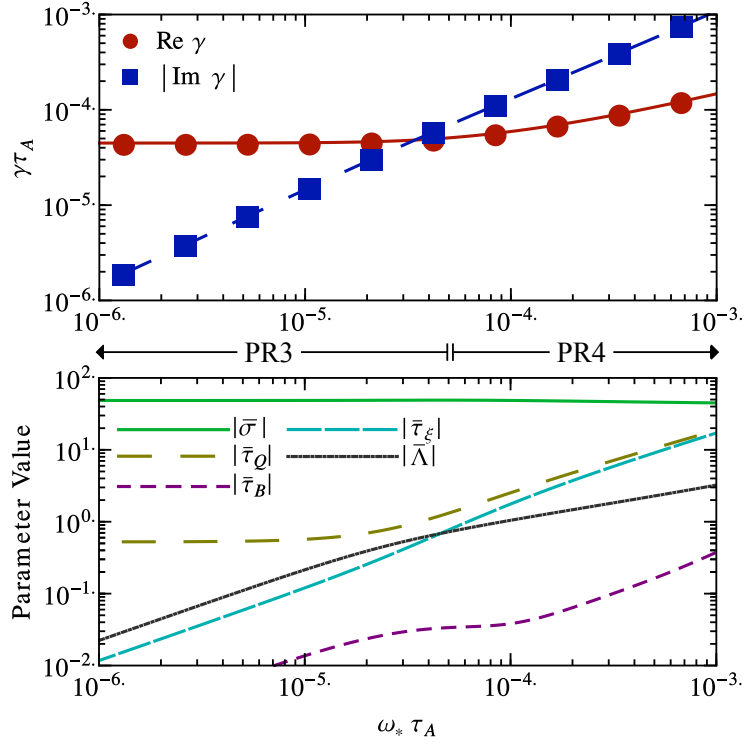}
\par\end{centering}

\caption{Scan C with $\mathbf{v}_{E\times B}=-\mathbf{v}_{\nabla p}$ and $f_{n}=1$.
The converged results from NIMROD runs (points) are compared with
the drift analytics from PR4 (lines, Eqn.~(\ref{eq:DispPR4})).\label{fig:scan3-4}}
\end{figure}

\begin{figure}
\begin{centering}
\includegraphics[width=0.45\paperwidth]{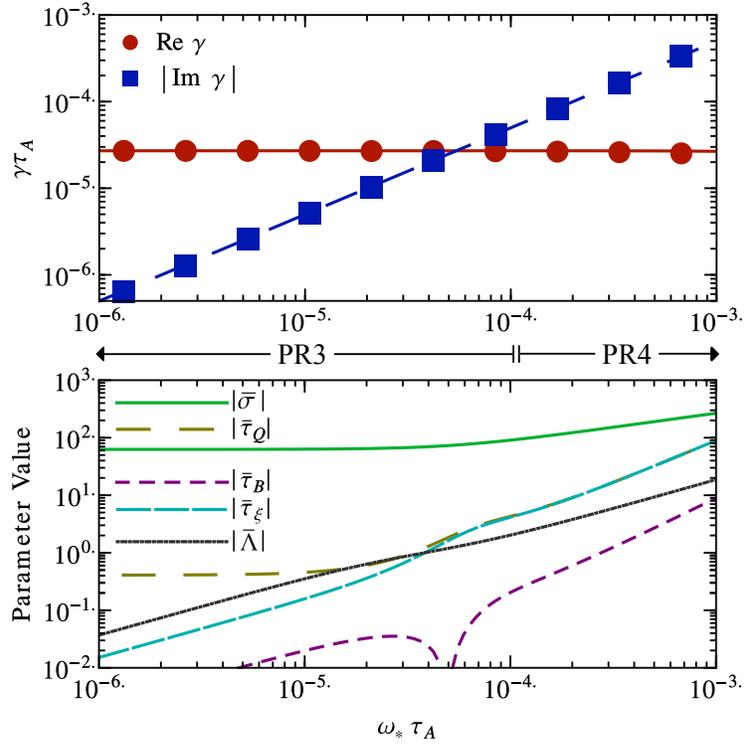}
\par\end{centering}

\caption{Scan C with a reduced electron mass, $\mu=2.7\times10^{-6}$, $\mathbf{v}_{E\times B}=0$
and $f_{n}=1$. The converged results from NIMROD runs (points) are
compared with the drift analytics from PR4 (lines, Eqn.~(\ref{eq:DispPR4})).\label{fig:scan3-5}}
\end{figure}

Figures \ref{fig:scan3-4} and \ref{fig:scan3-5} show the scan C
growth-rate comparisons at large $\hat{d}_{i}$ (2.048) and moderate
$\beta$ (0.1). These scans begin in PR3 and transition into PR4.
The NIMROD cases agree within 5\% and 1\% of the analytic results
for Figs.~\ref{fig:scan3-4} and \ref{fig:scan3-5}, respectively.
The cases in Fig.~\ref{fig:scan3-4} are essentially collisionless
and there is no drift stabilization as $\hat{\omega}_{*}$ increases,
instead the mode growth rate increases. In the collisionless regime
without the advective term in electron inertia, as currently implemented
in NIMROD, $S_{g}\rightarrow\hat{\gamma}^{-1}\hat{d}_{e}^{-2}$ and
Eqn.~(\ref{eq:DispPR3}) becomes
\begin{equation}
\hat{\gamma}_{e}^{2}\hat{\gamma}^{-1}=\hat{d}_{e}^{2}\hat{d}_{i}\hat{k}_{\parallel}^{\prime}\frac{\hat{\Delta^{\prime}}^{2}}{2\Gamma\left(\nicefrac{3}{4}\right)^{4}}\equiv\hat{\gamma}_{c}\;.
\end{equation}
In the limit of this equation where $\hat{\omega}_{*e}<<\hat{\gamma}_{c}$,
the mode grows at the drift-free growth rate and drifts at the electron
drift frequency, $\hat{\gamma}\simeq\hat{\gamma}_{c}+i\hat{\omega}_{*e}$.
In the limit where $\hat{\omega}_{*e}>>\hat{\gamma}_{c}$, the mode
grows proportionally to the square root of the drift frequency $\hat{\gamma}\simeq\left(\hat{\gamma}_{c}+\sqrt{\hat{\omega}_{*e}\hat{\gamma}_{c}}\right)/2+i\left(\hat{\omega}_{*e}+\sqrt{\hat{\omega}_{*e}\hat{\gamma}_{c}}/2\right)$.
This second limit explains the destabilization of the mode as seen
in the figure. It is of interest to note that if the advective term
is included in electron inertia then $S_{g}\rightarrow\hat{\gamma}_{e}^{-1}\hat{d}_{e}^{-2}$
and there is no growth rate increase. However, electron gyroviscosity
enters the equations on the same order and should also be retained.
The relevant physical effects within the boundary layer for these
near-collisionless cases illustrate the breakdown of the argument
to ignore electron advection and gyroviscosity presented in Sec.~\ref{sec:ModelEqnsOrdering}.
As the resonant condition causes the dominant terms in Ohm's law to
vanish, the boundary layer physics is determined by a balance of the
remaining, otherwise small terms. For the collisionless-drift-tearing
mode, these small terms include electron advection and gyroviscosity
(Ref.~\cite{Fitzpatrick_2010_PoP} includes these terms in PR5 without
drift effects). The cases in Fig.~\ref{fig:scan3-5} are identical
to those in Fig.~\ref{fig:scan3-4} except they are collisional through
the use of a small electron mass, $\mu=2.7\times10^{-6}$. For these
collisional cases the mode is not drift stabilized and simply rotates
with the electron fluid as predicted by Eqn.~(\ref{eq:DispPR3})
when $S_{g}\rightarrow v_{A}\mu_{0}/k_{\perp}\eta$.

Beyond the successful verification of the code in this electron-fluid-mediated
regime, the validity of the model remains in question. For first-order
electron-FLR model validity, one requires that $\rho_{e}/\delta=\sqrt{\beta\mu}\bar{\sigma}<<1$;
a condition that is satisfied for these cases. However, it is unlikely
that the simple electron-response model is sufficient to model the
collisionless dynamics of Fig.~\ref{fig:scan3-4}. Given that the
ion gyroviscous cancellation is incomplete (see Sec.~\ref{sub:IonGV}),
the implicit assumption in the model that $\nabla\cdot\mathbf{\Pi}_{e,gv}+m_{e}\mathbf{v}_{e}\cdot\nabla\mathbf{v}_{e}=0$
is likely not valid when $\hat{\omega}_{*}$ is large. Further study
and code development pertaining to this issue is required and outside
the scope of this work.

\begin{figure}
\begin{centering}
\includegraphics[width=0.45\paperwidth]{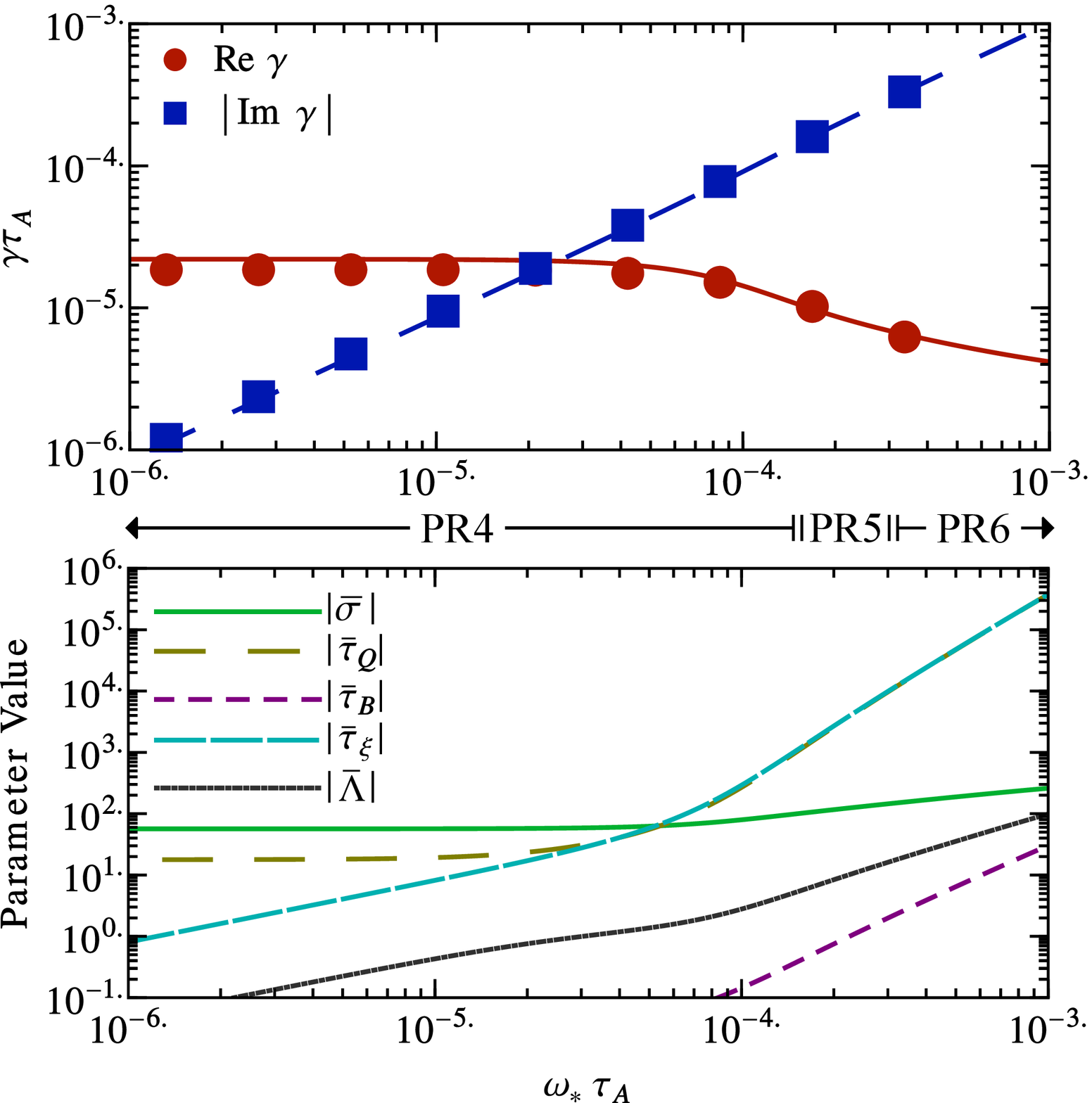}
\par\end{centering}

\caption{Scan D with $\mu=2.7\times10^{-6}$, $\mathbf{v}_{E\times B}=-\mathbf{v}_{\nabla p}$
and $f_{n}=0.6$. The converged results from NIMROD runs (points)
are compared with the drift analytics from PR4 (lines, Eqn.~(\ref{eq:DispPR4})).\label{fig:scan4-2}}
\end{figure}

The scan D verification exercise that begins in PR4 and transitions
through PR5 to PR6 at low $\beta$ ($1.56\times10^{-3}$) and large
$\hat{d}_{i}$ (2.048) is shown in Fig.~\ref{fig:scan4-2}. Although
we are not able to run a drift-verification scan while starting in
the semicollisional regime, PR5, this comparison does include cases
near this regime. In this regime the mode is weakly stabilized where
the growth rate is decreased by approximately a factor of five for
large values of $\omega_{*}/\gamma$. The discrepancy between the
analytics and the numerics for the first six cases is approximately
15\%, however, the right-most three cases, where the drift effects
are large, agree with the analytic theory within 7\%, 2\% and 0.2\%,
respectively.

\begin{figure}
\begin{centering}
\includegraphics[width=0.5\textwidth]{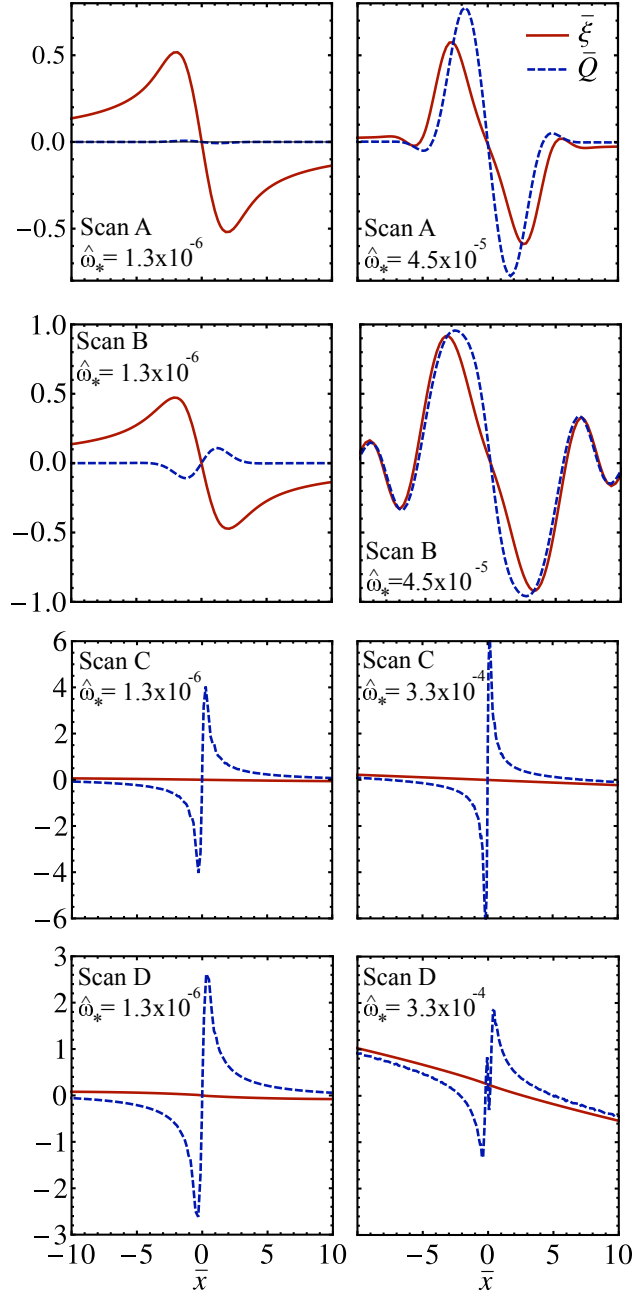}
\par\end{centering}

\caption{Eigenfunctions $\bar{\xi}$ and $\bar{Q}$ for scans A-D (top to bottom)
at small (left) and large (right) values of $\hat{\omega}_{*}$. The
plots for scan C correspond to Fig.~\ref{fig:scan3-5}\label{fig:eigen}}
\end{figure}

Figure \ref{fig:eigen} is a matrix of eigenfunction plots for scans
A-D at small and large values of $\hat{\omega}_{*}$. The scalings
of the $\bar{\xi}$ and $\bar{Q}$ are consistent with the assumptions
for the various regimes made in Sec.~\ref{sec:Dispersion}. For the
small-$\omega_{*}$, scan-A case $\bar{\xi}>>\bar{Q}$ which is reasonable
for a case near the single-fluid limit. When $\omega_{*}$ is large
(scan A and all of scan B), $\bar{\xi}\sim\bar{Q}$ in line with the
assumptions of PR2. For cases in PR4 through PR6 (scans C and D),
$\bar{Q}>>\bar{\xi}$, $\bar{Q}$ is larger than unity, and the eigenfunction
is more localized consistent with the orderings of $\bar{Q}\sim\bar{\sigma}^{1/2}\sim\bar{x}^{-1}$.
All cases except the large-$\omega_{*}$, scan-D case produce an odd
eigenfunction (only the odd component contributes to the growth rate,
a result of Eqn.~(\ref{eq:DispRelInt})). The large-$\omega_{*}$
scan D case has an even component which is in agreement with the discussion
of Secs.~\ref{sub:PR4} and \ref{sub:PR5} and large contributions
from $\bar{\tau}_{B}$. Finally, at large $\omega_{*}$ only scan
D has radial drift structures that extend to $\bar{x}=\pm100$ (not
shown). All other cases do not exhibit this structure and the eigenfunction
is highly localized within the resonant layer. Unlike previous computational
verification drift-tearing work \cite{Biskamp_78_NF_1059}, our computations
do not exhibit significant influence from the computational boundary
condition.

\section{Concluding Discussion \label{sec:Conclusions}}

\noindent This work is both an analytic and computational investigation
of drift tearing with an unreduced, extended-MHD model. Our new analytic
results have been used to verify the implementation of the extended-MHD
equations within the NIMROD code. As the tearing-layer dynamics result
from the balance of otherwise small terms, this verification is a
novel way to test the extended-MHD implementation. Our new analytic
results describe the experimentally relevant portion of the drift-tearing
phase space. Within this phase space, there is the potential for varying
degrees of drift stabilization: there is a weakly stabilizing effect
at either small $d_{i}$ and moderate $\beta$ or at large $d_{i}$
and small $\beta$, complete stabilization is possible at moderate
$d_{i}$ and $\beta$ and there is no stabilization at large $d_{i}$
and moderate $\beta$ where the ion dynamics are decoupled from the
mode. We emphasize that our definition of moderate $\beta$ encompasses
the values that are pertinent for a fusion reactor ($\beta\sim1\%-25\%$).
There are some caveats to the applicability of this work when one
considers the validity of the first-order ion FLR model. However,
we argue that this model may still be qualitatively valid when the
ion gyroradius is no longer small, as the mode transitions to one
dominated solely by the electron-fluid dynamics (given a sufficient
electron-dynamics model). Our results can not be directly applied
tokamak discharges, as we do not retain the effects of ion gyroviscosity
and plasma shaping and curvature. Instead, the ultimate benefit of
this work is to provide enhanced confidence in nonlinear, extended-MHD,
boundary-layer-dynamics computations of tokamak discharges with reconstructed
profiles and realistic geometry.

\begin{acknowledgments}
The authors would like to thank Carl Sovinec and Chris Hegna for stimulating
discussions. This work is supported by US Department of Energy grants
DE-FC02-06ER54875 and DE-FG02-08ER54972. This research used resources
of the National Energy Research Scientific Computing Center, which
is supported by the Office of Science of the U.S. Department of Energy
under Contract No. DE-AC02-05CH11231.
\end{acknowledgments}
\bibliographystyle{apsrev4-1}
\bibliography{index}

\begin{thebibliography}{19}%
\makeatletter
\providecommand \@ifxundefined [1]{%
 \@ifx{#1\undefined}
}%
\providecommand \@ifnum [1]{%
 \ifnum #1\expandafter \@firstoftwo
 \else \expandafter \@secondoftwo
 \fi
}%
\providecommand \@ifx [1]{%
 \ifx #1\expandafter \@firstoftwo
 \else \expandafter \@secondoftwo
 \fi
}%
\providecommand \natexlab [1]{#1}%
\providecommand \enquote  [1]{``#1''}%
\providecommand \bibnamefont  [1]{#1}%
\providecommand \bibfnamefont [1]{#1}%
\providecommand \citenamefont [1]{#1}%
\providecommand \href@noop [0]{\@secondoftwo}%
\providecommand \href [0]{\begingroup \@sanitize@url \@href}%
\providecommand \@href[1]{\@@startlink{#1}\@@href}%
\providecommand \@@href[1]{\endgroup#1\@@endlink}%
\providecommand \@sanitize@url [0]{\catcode `\\12\catcode `\$12\catcode
  `\&12\catcode `\#12\catcode `\^12\catcode `\_12\catcode `\%12\relax}%
\providecommand \@@startlink[1]{}%
\providecommand \@@endlink[0]{}%
\providecommand \url  [0]{\begingroup\@sanitize@url \@url }%
\providecommand \@url [1]{\endgroup\@href {#1}{\urlprefix }}%
\providecommand \urlprefix  [0]{URL }%
\providecommand \Eprint [0]{\href }%
\providecommand \doibase [0]{http://dx.doi.org/}%
\providecommand \selectlanguage [0]{\@gobble}%
\providecommand \bibinfo  [0]{\@secondoftwo}%
\providecommand \bibfield  [0]{\@secondoftwo}%
\providecommand \translation [1]{[#1]}%
\providecommand \BibitemOpen [0]{}%
\providecommand \bibitemStop [0]{}%
\providecommand \bibitemNoStop [0]{.\EOS\space}%
\providecommand \EOS [0]{\spacefactor3000\relax}%
\providecommand \BibitemShut  [1]{\csname bibitem#1\endcsname}%
\let\auto@bib@innerbib\@empty
\bibitem [{\citenamefont {Furth}\ \emph {et~al.}(1963)\citenamefont {Furth},
  \citenamefont {Killeen},\ and\ \citenamefont
  {Rosenbluth}}]{Furth_63_PoF_459}%
  \BibitemOpen
  \bibfield  {author} {\bibinfo {author} {\bibfnamefont {H.~P.}\ \bibnamefont
  {Furth}}, \bibinfo {author} {\bibfnamefont {J.}~\bibnamefont {Killeen}}, \
  and\ \bibinfo {author} {\bibfnamefont {M.~N.}\ \bibnamefont {Rosenbluth}},\
  }\href {\doibase DOI:10.1063/1.1706761} {\bibfield  {journal} {\bibinfo
  {journal} {Physics of Fluids}\ }\textbf {\bibinfo {volume} {6}},\ \bibinfo
  {pages} {459} (\bibinfo {year} {1963})}\BibitemShut {NoStop}%
\bibitem [{\citenamefont {Braginskii}(1965)}]{Braginskii_65__}%
  \BibitemOpen
  \bibfield  {author} {\bibinfo {author} {\bibfnamefont {S.~I.}\ \bibnamefont
  {Braginskii}},\ }\href@noop {} {\emph {\bibinfo {title} {Transport Properties
  in a Plasma in Review of Plasma Physics}}},\ edited by\ \bibinfo {editor}
  {\bibfnamefont {M.~A.}\ \bibnamefont {Leontovich}},\ Vol.~\bibinfo {volume}
  {1}\ (\bibinfo  {publisher} {Consultants Bureau},\ \bibinfo {address} {New
  York},\ \bibinfo {year} {1965})\BibitemShut {NoStop}%
\bibitem [{\citenamefont {Catto}\ and\ \citenamefont
  {Simakov}(2004)}]{Catto_04_PoP_90}%
  \BibitemOpen
  \bibfield  {author} {\bibinfo {author} {\bibfnamefont {P.~J.}\ \bibnamefont
  {Catto}}\ and\ \bibinfo {author} {\bibfnamefont {A.~N.}\ \bibnamefont
  {Simakov}},\ }\href {\doibase DOI:10.1063/1.1632496} {\bibfield  {journal}
  {\bibinfo  {journal} {Physics of Plasmas}\ }\textbf {\bibinfo {volume}
  {11}},\ \bibinfo {pages} {90} (\bibinfo {year} {2004})}\BibitemShut {NoStop}%
\bibitem [{\citenamefont {Ramos}(2010)}]{Ramos_2010_PoP1}%
  \BibitemOpen
  \bibfield  {author} {\bibinfo {author} {\bibfnamefont {J.~J.}\ \bibnamefont
  {Ramos}},\ }\href {\doibase http://dx.doi.org/10.1063/1.3454368} {\bibfield
  {journal} {\bibinfo  {journal} {Physics of Plasmas (1994-present)}\ }\textbf
  {\bibinfo {volume} {17}},\ \bibinfo {eid} {082502} (\bibinfo {year}
  {2010})}\BibitemShut {NoStop}%
\bibitem [{\citenamefont {Ramos}(2011)}]{Ramos_2011_PoP1}%
  \BibitemOpen
  \bibfield  {author} {\bibinfo {author} {\bibfnamefont {J.~J.}\ \bibnamefont
  {Ramos}},\ }\href {\doibase http://dx.doi.org/10.1063/1.3647568} {\bibfield
  {journal} {\bibinfo  {journal} {Physics of Plasmas (1994-present)}\ }\textbf
  {\bibinfo {volume} {18}},\ \bibinfo {eid} {102506} (\bibinfo {year}
  {2011})}\BibitemShut {NoStop}%
\bibitem [{\citenamefont {Coppi}(1964)}]{Coppi_64_PoF_1501}%
  \BibitemOpen
  \bibfield  {author} {\bibinfo {author} {\bibfnamefont {B.}~\bibnamefont
  {Coppi}},\ }\href {\doibase DOI:10.1063/1.1711405} {\bibfield  {journal}
  {\bibinfo  {journal} {Physics of Fluids}\ }\textbf {\bibinfo {volume} {7}},\
  \bibinfo {pages} {1501} (\bibinfo {year} {1964})}\BibitemShut {NoStop}%
\bibitem [{\citenamefont {King}\ \emph {et~al.}(2011)\citenamefont {King},
  \citenamefont {Sovinec},\ and\ \citenamefont {Mirnov}}]{King_11_PoP_42303}%
  \BibitemOpen
  \bibfield  {author} {\bibinfo {author} {\bibfnamefont {J.~R.}\ \bibnamefont
  {King}}, \bibinfo {author} {\bibfnamefont {C.~R.}\ \bibnamefont {Sovinec}}, \
  and\ \bibinfo {author} {\bibfnamefont {V.~V.}\ \bibnamefont {Mirnov}},\
  }\href {\doibase DOI:10.1063/1.3571599} {\bibfield  {journal} {\bibinfo
  {journal} {Physics of Plasmas}\ }\textbf {\bibinfo {volume} {18}},\ \bibinfo
  {pages} {042303} (\bibinfo {year} {2011})}\BibitemShut {NoStop}%
\bibitem [{\citenamefont {Ahedo}\ and\ \citenamefont
  {Ramos}(2009)}]{Ahedo_09_PPaCF_55018}%
  \BibitemOpen
  \bibfield  {author} {\bibinfo {author} {\bibfnamefont {E.}~\bibnamefont
  {Ahedo}}\ and\ \bibinfo {author} {\bibfnamefont {J.~J.}\ \bibnamefont
  {Ramos}},\ }\href {http://stacks.iop.org/0741-3335/51/i=5/a=055018}
  {\bibfield  {journal} {\bibinfo  {journal} {Plasma Physics and Controlled
  Fusion}\ }\textbf {\bibinfo {volume} {51}},\ \bibinfo {pages} {055018}
  (\bibinfo {year} {2009})}\BibitemShut {NoStop}%
\bibitem [{\citenamefont {Drake}\ and\ \citenamefont
  {Lee}(1977)}]{Drake_77_PoF_1341}%
  \BibitemOpen
  \bibfield  {author} {\bibinfo {author} {\bibfnamefont {J.~F.}\ \bibnamefont
  {Drake}}\ and\ \bibinfo {author} {\bibfnamefont {Y.~C.}\ \bibnamefont
  {Lee}},\ }\href {\doibase DOI:10.1063/1.862017} {\bibfield  {journal}
  {\bibinfo  {journal} {Physics of Fluids}\ }\textbf {\bibinfo {volume} {20}},\
  \bibinfo {pages} {1341} (\bibinfo {year} {1977})}\BibitemShut {NoStop}%
\bibitem [{\citenamefont {Bulanov}\ \emph {et~al.}(1992)\citenamefont
  {Bulanov}, \citenamefont {Pegoraro},\ and\ \citenamefont
  {Sakharov}}]{Bulanov_92_PoFB_2499}%
  \BibitemOpen
  \bibfield  {author} {\bibinfo {author} {\bibfnamefont {S.~V.}\ \bibnamefont
  {Bulanov}}, \bibinfo {author} {\bibfnamefont {F.}~\bibnamefont {Pegoraro}}, \
  and\ \bibinfo {author} {\bibfnamefont {A.~S.}\ \bibnamefont {Sakharov}},\
  }\href {\doibase DOI:10.1063/1.860467} {\bibfield  {journal} {\bibinfo
  {journal} {Physics of Fluids B}\ }\textbf {\bibinfo {volume} {4}},\ \bibinfo
  {pages} {2499} (\bibinfo {year} {1992})}\BibitemShut {NoStop}%
\bibitem [{\citenamefont {Mirnov}\ \emph {et~al.}(2004)\citenamefont {Mirnov},
  \citenamefont {Hegna},\ and\ \citenamefont {Prager}}]{Mirnov_04_PoP_4468}%
  \BibitemOpen
  \bibfield  {author} {\bibinfo {author} {\bibfnamefont {V.~V.}\ \bibnamefont
  {Mirnov}}, \bibinfo {author} {\bibfnamefont {C.~C.}\ \bibnamefont {Hegna}}, \
  and\ \bibinfo {author} {\bibfnamefont {S.~C.}\ \bibnamefont {Prager}},\
  }\href {\doibase DOI:10.1063/1.1773778} {\bibfield  {journal} {\bibinfo
  {journal} {Physics of Plasmas}\ }\textbf {\bibinfo {volume} {11}},\ \bibinfo
  {pages} {4468} (\bibinfo {year} {2004})}\BibitemShut {NoStop}%
\bibitem [{\citenamefont {Biskamp}(1978)}]{Biskamp_78_NF_1059}%
  \BibitemOpen
  \bibfield  {author} {\bibinfo {author} {\bibfnamefont {D.}~\bibnamefont
  {Biskamp}},\ }\href {http://stacks.iop.org/0029-5515/18/i=8/a=003} {\bibfield
   {journal} {\bibinfo  {journal} {Nuclear Fusion}\ }\textbf {\bibinfo {volume}
  {18}},\ \bibinfo {pages} {1059} (\bibinfo {year} {1978})}\BibitemShut
  {NoStop}%
\bibitem [{\citenamefont {Sovinec}\ \emph {et~al.}(2004)\citenamefont
  {Sovinec}, \citenamefont {Glasser}, \citenamefont {Gianakon}, \citenamefont
  {Barnes}, \citenamefont {Nebel}, \citenamefont {Kruger}, \citenamefont
  {Schnack}, \citenamefont {Plimpton}, \citenamefont {Tarditi},\ and\
  \citenamefont {Chu}}]{Sovinec_04_JoCP_355}%
  \BibitemOpen
  \bibfield  {author} {\bibinfo {author} {\bibfnamefont {C.}~\bibnamefont
  {Sovinec}}, \bibinfo {author} {\bibfnamefont {A.}~\bibnamefont {Glasser}},
  \bibinfo {author} {\bibfnamefont {T.}~\bibnamefont {Gianakon}}, \bibinfo
  {author} {\bibfnamefont {D.}~\bibnamefont {Barnes}}, \bibinfo {author}
  {\bibfnamefont {R.}~\bibnamefont {Nebel}}, \bibinfo {author} {\bibfnamefont
  {S.}~\bibnamefont {Kruger}}, \bibinfo {author} {\bibfnamefont
  {D.}~\bibnamefont {Schnack}}, \bibinfo {author} {\bibfnamefont
  {S.}~\bibnamefont {Plimpton}}, \bibinfo {author} {\bibfnamefont
  {A.}~\bibnamefont {Tarditi}}, \ and\ \bibinfo {author} {\bibfnamefont
  {M.}~\bibnamefont {Chu}},\ }\href {\doibase 10.1016/j.jcp.2003.10.004}
  {\bibfield  {journal} {\bibinfo  {journal} {Journal of Computational
  Physics}\ }\textbf {\bibinfo {volume} {195}},\ \bibinfo {pages} {355 }
  (\bibinfo {year} {2004})}\BibitemShut {NoStop}%
\bibitem [{\citenamefont {Sovinec}\ and\ \citenamefont
  {King}(2010)}]{Sovinec_10_JoCP_5803}%
  \BibitemOpen
  \bibfield  {author} {\bibinfo {author} {\bibfnamefont {C.}~\bibnamefont
  {Sovinec}}\ and\ \bibinfo {author} {\bibfnamefont {J.}~\bibnamefont {King}},\
  }\href {\doibase 10.1016/j.jcp.2010.04.022} {\bibfield  {journal} {\bibinfo
  {journal} {Journal of Computational Physics}\ }\textbf {\bibinfo {volume}
  {229}},\ \bibinfo {pages} {5803 } (\bibinfo {year} {2010})}\BibitemShut
  {NoStop}%
\bibitem [{\citenamefont {Fitzpatrick}(2010)}]{Fitzpatrick_2010_PoP}%
  \BibitemOpen
  \bibfield  {author} {\bibinfo {author} {\bibfnamefont {R.}~\bibnamefont
  {Fitzpatrick}},\ }\href {\doibase 10.1063/1.3374427} {\bibfield  {journal}
  {\bibinfo  {journal} {Physics of Plasmas}\ }\textbf {\bibinfo {volume}
  {17}},\ \bibinfo {eid} {042101} (\bibinfo {year} {2010})}\BibitemShut
  {NoStop}%
\bibitem [{\citenamefont {Ramos}(2005)}]{Ramos_05_PoP_112301}%
  \BibitemOpen
  \bibfield  {author} {\bibinfo {author} {\bibfnamefont {J.~J.}\ \bibnamefont
  {Ramos}},\ }\href {\doibase DOI:10.1063/1.2114747} {\bibfield  {journal}
  {\bibinfo  {journal} {Physics of Plasmas}\ }\textbf {\bibinfo {volume}
  {12}},\ \bibinfo {pages} {112301} (\bibinfo {year} {2005})}\BibitemShut
  {NoStop}%
\bibitem [{\citenamefont {Ahedo}\ and\ \citenamefont
  {Ramos}(2012)}]{Ahedo_2012_PoP1}%
  \BibitemOpen
  \bibfield  {author} {\bibinfo {author} {\bibfnamefont {E.}~\bibnamefont
  {Ahedo}}\ and\ \bibinfo {author} {\bibfnamefont {J.~J.}\ \bibnamefont
  {Ramos}},\ }\href {\doibase http://dx.doi.org/10.1063/1.4739787} {\bibfield
  {journal} {\bibinfo  {journal} {Physics of Plasmas (1994-present)}\ }\textbf
  {\bibinfo {volume} {19}},\ \bibinfo {eid} {072519} (\bibinfo {year}
  {2012})}\BibitemShut {NoStop}%
\bibitem [{\citenamefont {Hazeltine}\ and\ \citenamefont
  {Meiss}(2013)}]{hazeltineMeiss}%
  \BibitemOpen
  \bibfield  {author} {\bibinfo {author} {\bibfnamefont {R.}~\bibnamefont
  {Hazeltine}}\ and\ \bibinfo {author} {\bibfnamefont {J.}~\bibnamefont
  {Meiss}},\ }\href {http://books.google.com/books?id=LDPDAgAAQBAJ} {\emph
  {\bibinfo {title} {Plasma Confinement}}},\ Dover Books on Physics\ (\bibinfo
  {publisher} {Dover Publications},\ \bibinfo {year} {2013})\BibitemShut
  {NoStop}%
\bibitem [{\citenamefont {Schnack}\ \emph {et~al.}(2013)\citenamefont
  {Schnack}, \citenamefont {Cheng}, \citenamefont {Barnes},\ and\ \citenamefont
  {Parker}}]{Schnack_13_PoP}%
  \BibitemOpen
  \bibfield  {author} {\bibinfo {author} {\bibfnamefont {D.~D.}\ \bibnamefont
  {Schnack}}, \bibinfo {author} {\bibfnamefont {J.}~\bibnamefont {Cheng}},
  \bibinfo {author} {\bibfnamefont {D.~C.}\ \bibnamefont {Barnes}}, \ and\
  \bibinfo {author} {\bibfnamefont {S.~E.}\ \bibnamefont {Parker}},\ }\href
  {\doibase http://dx.doi.org/10.1063/1.4811468} {\bibfield  {journal}
  {\bibinfo  {journal} {Physics of Plasmas (1994-present)}\ }\textbf {\bibinfo
  {volume} {20}},\ \bibinfo {eid} {062106} (\bibinfo {year}
  {2013})}\BibitemShut {NoStop}%
\end{thebibliography}%

\end{document}